\documentclass[bibyear,breaklinks=true]{aa}

\usepackage{silence}
\usepackage{verbatim}
\usepackage{graphicx}
\usepackage{lscape}
\usepackage{txfonts}
\usepackage{stackengine}
\usepackage{subcaption}
\usepackage{geometry}
\usepackage{float}
\usepackage{tabularx}
\usepackage{longtable}
\usepackage{booktabs}
\usepackage{gensymb}
\usepackage{textgreek}
\usepackage{textcomp}
\usepackage{orcidlink}

\usepackage{natbib,twoopt}

\bibpunct{(}{)}{;}{a}{}{,}
\hypersetup{
colorlinks=true,
linkcolor=red, 
citecolor=blue, 
urlcolor=blue 
}

\begin{document}

\title{Repeating flares, X-ray outbursts and delayed infrared emission:\\A comprehensive compilation of optical tidal disruption events}

\subtitle{TDECat}

\author{D.A. Langis \inst{1,2}\orcidlink{0009-0003-9365-9073}
	\and
	I. Liodakis\inst{1,3}\orcidlink{0000-0001-9200-4006}
	\and
	K.I.I. Koljonen\inst{4}\orcidlink{0000-0002-9677-1533}
	\and
	A. Paggi\inst{1}\orcidlink{0000-0002-5646-2410}
	\and
	N. Globus\inst{5,6}\orcidlink{0000-0001-9011-0737}
	\and
	L. Wyrzykowski\inst{7,8}\orcidlink{0000-0002-9658-6151}
	\and
	P.J. Miko{\l}ajczyk \inst{7,8,9} \orcidlink{0000-0001-8916-8050}
	\and 
	K. Kotysz \inst{7,9} \orcidlink{0000-0003-4960-7463}
	\and
	P. Zieli{\'n}ski \inst{10} \orcidlink{0000-0001-6434-9429}
	\and
	N. Ihanec\inst{11,7}
	\and
	J. Ding\inst{12,13}\orcidlink{0009-0007-1206-8239}
	\and
	D. Morshed\inst{14}
	\and
	Z. Torres\inst{15}
}
\titlerunning{TDECat}
\authorrunning{D.A. Langis}
\institute{Institute of Astrophysics, FORTH, GR-71110 Heraklion, Greece  \email{dlangis@physics.uoc.gr}
	\and 
	Department of Physics, University of Crete, 71003 Heraklion, Greece
	\and
	NASA Marshall Space Flight Center, Huntsville, AL 35812, USA
	\and
	Institutt for Fysikk, Norwegian University of Science and Technology, Høgskloreringen 5, Trondheim 7491, Norway
	\and
	Instituto de Astronomía, Universidad Nacional Autónoma de México Campus Ensenada, Carr. Tijuana-Ensenada km107, Ensenada, BC 22800, México
	\and
	Kavli Institute for Particle Astrophysics and Cosmology, Stanford University, Stanford, CA 94305, USA
	\and
	Astronomical Observatory, University of Warsaw, Al. Ujazdowskie 4, 00-478 Warszawa, Poland
	\and
	Astrophysics Division, National Centre for Nuclear Research, Pasteura 7, 02-093, Warsaw, Poland 
	\and
	Astronomical Institute, University of Wrocław, Kopernika 11, 51-622 Wrocław, Poland
	\and
	Institute of Astronomy, Faculty of Physics, Astronomy and Informatics, Nicolaus Copernicus University, Grudziądzka 5, 87-100 Toruń, Poland
	\and
	Isaac Newton Group of Telescopes, Apartado 321, E-38700 Santa Cruz de la Palma, Spain
	\and
	Princeton University, Princeton, NJ 08544, USA
	\and
	Rye Country Day School, NY 10580, USA
	\and 
	Head-Royce School, CA 94602, USA
	\and
	Harbor High School, CA 95062, USA
}

\date{Received ; accepted }

\abstract
{Tidal disruption events (TDEs) have been proposed as valuable laboratories for studying dormant black holes. However, progress in this field has been hampered by the limited number of observed events. In this work, we present TDECat, a comprehensive catalogue of 134 confirmed TDEs (131 optical TDEs and three jetted TDEs) discovered up to the end of 2024, accompanied by multi-wavelength photometry (X-ray, UV, optical, and infrared) and publicly available spectra. We also study the statistical properties, spectral classifications, and multi-band variability of these events. Using a Bayesian Blocks algorithm, we determined the duration, rise time ($t_{rise}$), decay time ($t_{decay}$), and their ratio for 103 flares in our sample. We find that these timescales follow a log-normal distribution. Furthermore, our spectral analysis shows that most optical TDEs belong to the TDE-H+He class, followed by the TDE-H, TDE-He, and TDE-featureless classes, which is consistent with expectations from main-sequence star disruption. Using archival observations, we identified three new potentially repeating TDEs, namely, AT 2024pvu, AT 2022exr, and AT 2021uvz, increasing the number of known repeating events. In both newly identified and previously known cases, the secondary flares exhibit a similar shape to the primary. We also examined the infrared and X-ray emission from the TDEs in our catalogue, and find that 14 out of the 18 infrared events have associated X-ray emission, strongly suggesting a potential correlation. Finally, we find that for three sub-samples (repeating flares, infrared-emitting events, and X-ray-emitting events), the spectral classes are unlikely to be randomly distributed, suggesting a connection between spectral characteristics and multi-wavelength emission. TDEcat enables large-scale population studies across wavelengths and spectral classes, providing essential tools for navigating the data-rich era of upcoming surveys such as the Legacy Survey of Space and Time.}

\maketitle
\nolinenumbers

\section{Introduction}\label{sect:Intro}

A tidal disruption event (TDE) occurs when a star ventures too close to a supermassive black hole (SMBH), where differential gravitational forces exceed the star's self gravity, causing it to be torn apart by tidal forces. This occurs only if the tidal radius, $R_T$, which is the critical radius within which an object is disrupted by the SMBH, is larger than the event horizon. There is a theoretical upper limit to the SMBH's mass known as the Hills mass, since $R_T \propto M^{1/3}_{BH}$. For a solar mass star, this limit is approximately $10^8 M_\odot$ \citep{Hills1975hillsmass}. About half of the stellar material remains bound to the SMBH forming an accretion disc, while the other half is ejected \citep{Rees1988tde}. This process produces a characteristic flare with a sharp rise in brightness, followed by a gradual decay leading to a plateau. These flares can be observed across the electromagnetic spectrum, from radio to hard X-rays and typically exhibit broad hydrogen and helium lines. The expected detection rate of optical TDEs has been estimated at a few $10^{-5}$ galaxy$^{-1}$ yr$^{-1}$ \citep{Yao2023sample}, which is at least around one order of magnitude smaller than the theoretically derived rates for TDEs \citep[e.g. ][]{Magorrian1999MNRASTheoreticalRates,Wang2004ApJTheoreticalRates,Stone2016MNRAScitationsREF}. Over the past half decade, approximately 10-20 new optical TDEs have been detected annually.

The first TDE was identified during the ROSAT all-sky X-ray survey, with the observed soft X-ray emission attributed to a newly formed accretion disc \citep{Bade1996X-rayTDE, Grupe1999X-rayTDE, Saxton2020X-rays}. Since then, numerous TDEs have been detected by various instruments and surveys. In the X-ray band, both XMM-Newton \citep[e.g. ][]{Esquej2007SDSSJ1323, Lin2011ApJXMMTDE, Saxton2019A&AXMMTDE} and eROSITA \citep[e.g. ][]{Saxton2020X-rays, Sazonov2021XrayTDE, Grotova2025A&AeROSITATDEs} have observed multiple TDEs and candidates. The Chandra X-ray Observatory and the BAT telescope aboard the Neil Gehrels \textit{Swift} Observatory \citep{2005SSRv..120...95R, Burrows2005SwiftXray} have played important roles in follow-up X-ray observations of TDEs discovered at optical wavelengths. Additionally, \textit{Swift}'s UVOT telescope has enabled observations of these events in the ultraviolet.

Optical surveys have significantly contributed to our understanding of TDEs. The All-Sky Automated Survey for Supernovae \citep[ASAS-SN; ][]{Shappee2014ASASSNnotsure, Kochanek2017ASASSN, Hart2023ASASSNskyPatrol}, a global network of small automated telescopes scanning the entire sky for supernovae and other transients, has identified numerous TDEs through their distinctive optical flares \citep[e.g. ][]{Holoien2014ASASSN14ae, Holoien2016ASASSN14liOptXray}. Similarly, the Zwicky Transient Facility \citep[ZTF; ][]{Graham2019ZTFScObj, Bellm2019ZTFSys, Masci2019ZTFDataPr}, with its high-cadence, wide-field survey capabilities, has played a key role in the discovery of optical TDEs \citep[e.g. ][]{vanVelzen2021sample, Hammerstein2023ZTF1}, by rapidly surveying vast areas of the sky. The Asteroid Terrestrial-impact Last Alert System \citep[ATLAS; ][]{Tonry2018ATLAS,Heinze2018ATLAS,Smith2020ATLAS,Shingles2021ATLAS} has also been crucial to long-term monitoring of TDEs \citep[e.g. ][]{Nicholl2020MNRASATLASTDE,Earl2025ApJAT2020nov} by providing photometric data in two optical wavebands.

The Gaia photometric Science Alerts system \citep[GSA\footnote{\url{http://gsaweb.ast.cam.ac.uk/alerts}};][]{Hodgkin2021GSA} has also contributed to TDE discoveries by detecting and providing optical photometry of transient events \citep[e.g. ][]{Wevers2022A&AGAIATDE}. Several other surveys were instrumental in early TDE observations, including the Catalina Real-Time Transient Survey \citep[CRTS; ][]{Drake2009CRTS}, Lincoln Near-Earth Asteroid Research \citep[LINEAR; ][]{Stokes2000LINEAR}, Panoramic Survey Telescope and Rapid Response System \citep[Pan-STARRS; ][]{Chambers2016PS1,Waters2020PS1,Magnier2020PS1,Magnier2020PS1_,Magnier2020PS1__,Flewelling2020PS1}, the Palomar Transient Factory \citep[PTF;][]{Law2009PTF,Rau2009PTF,Arcavi2014PTFTDEs} and its successor, the Intermediate Palomar Transient Factory \citep[iPTF\footnote{\url{https://www.ptf.caltech.edu/iptf/}}; ][]{Hung2017iPTF16axa,Blagorodnova2017ApJiPTF16fnl, Blagorodnova2019iPTF15af}.

Beyond optical wavelengths, TDEs have been observed in the infrared, initially by the Wide-field Infrared Survey Explorer \citep[WISE; ][]{Wright2010WISE} and later by the Near-Earth Object WISE Reactivation \citep[NEOWISE; ][]{Mainzer2011NEOWISE,Mainzer2014NEOWISE, Jiang2016ASASSN14liIR, Dou2016SDSSJ0952}. Spectroscopic follow-up observations, including the Sloan Digital Sky Survey \citep[SDSS; ][]{York2000SDSS, vanVelzen2011ApJSDSS} and the Palomar 200-inch Hale Telescope, have been crucial in classifying TDEs based on their distinctive spectral signatures.

Despite the growing numbers of TDEs no comprehensive catalogue currently exists. Such a resource would enable robust statistical analyses by providing a dataset large enough to conduct meaningful studies, even if the sample is incomplete. The primary goal of this paper is to compile and present all publicly available photometric and spectroscopic data for confirmed optical TDEs up to the end of 2024. Additionally, we investigate different sub-categories of optical TDEs, including those displaying repeating flares, X-ray outbursts, and IR emission.

Our paper is organised as follows: In Sect. \ref{sect:Collection}, we outline the construction of our main sample of confirmed optical TDEs, along with a sample of TDE candidates. In Sect. \ref{sec:data} we describe the data collection process for the catalogue, while in Sect. \ref{sect:Explore} we delve into the properties of our sample. Finally, in Sect. \ref{sect:Discussion} we discuss our results and their implications, and we follow it with a summary in Sect. \ref{sect:Conclusions}.

\section{Sample selection}\label{sect:Collection}

Our sample consists of events identified either in the Transient Name Server (TNS\footnote{\url{https://www.wis-tns.org/}}; 105 sources) or in the literature (29 sources). We differentiate between confirmed TDEs and TDE candidates by organising them into two separate catalogues. The main catalogue consists of TDEs that can be found at least in one optical photometric survey and have an available classification spectrum from the time of the flare. We also include TDEs with featureless spectra (see Sect. \ref{subsect:Spectral Type}) and the three widely accepted on-axis jetted TDEs from the literature. TDEs that do not meet these criteria are designated as TDE candidates and are included in a separate TDE candidates catalogue.

\subsection{Main catalogue}\label{subsec:Main}

The first step in constructing our sample was to retrieve all objects classified as TDEs from TNS. At the time of this study, TNS listed 98 classified TDEs up to the end of 2024. However, AT 2018meh is the same event as AT 2023clx, so we include only AT 2023clx, reducing the TNS-TDE count to 97. From TNS, we also include 4 additional TDE-H+He events, 3 TDE-He events (see Sect. \ref{subsect:Spectral Type} for spectral classification) and a TDE-H+He event that is classified in an AstroNote (AT 2024ule\footnote{\url{https://www.wis-tns.org/astronotes/astronote/2024-318}}). Hence, our TNS-TDE sample consists of a total of 105 transients.

Beyond TNS, additional TDEs have been identified in various published sample studies. \citet{Hammerstein2023ZTF1} present a sample of 30 TDEs, observed during the first phase of the ZTF survey (see their Table 1). Of these, 12 are not classified in TNS: AT 2018lni, AT 2018jbv, AT 2019cho, AT 2019mha, AT 2019meg, AT 2020ddv, AT 2020ocn, AT 2020opy, AT 2020mbq, AT 2020qhs, AT 2020riz and AT 2020ysg. We note that this sample includes almost all the TDEs from the sample of \citet{vanVelzen2021sample}, except for AT 2019eve. This transient showed spectral and light curve evolution that made its initial TDE classification ambiguous \citep{Hammerstein2023ZTF1}. For this reason, we opt to include AT 2019eve in Sect. \ref{subsec:Honorable}, where we present strong TDE candidates.

Another TDE sample is presented in \citet{Yao2023sample} which lists 33 TDEs (see their Table 3). Of these, only three are absent from both the \cite{Hammerstein2023ZTF1} sample and the TNS-TDE sample, namely, AT 2019baf, AT 2019cmw and AT 2020abri. While these three were classified as TDEs in \citet{Yao2023sample}, there are uncertainties regarding their classification. AT 2019baf also appears in \citet{Somalwar2023AT2019baf} as a TDE, but it lacks a classification spectrum from the time of the flare. Additionally, \citet{Somalwar2023AT2019baf} cite an unpublished work as the basis for its classification. A similar situation applies to AT 2019cmw, which is also referenced in an unpublished paper in \citet{Yao2023sample}. AT 2020abri, meanwhile, has an optical spectrum taken 395 days after the peak of its optical flare, well after the event had likely faded. Its classification is based on 1) persistent blue colour and lack of cooling, which is inconsistent with the majority of supernoave and 2) the combination of weak H$\alpha$ emission and strong H$\delta$ absorption, which indicate that the host galaxy is post-starburst (where TDE rates are enhanced; see Sect. \ref{subsect:Repeating}). Since none of these three sources have spectra from the time of their flares, we exclude them from the main sample and place them in the TDE candidates sample.

Additionally, several studies focus on individual TDEs that are neither classified in TNS nor included in large sample studies \citep[e.g.][]{vanVelzen2021sample, Hammerstein2023ZTF1, Yao2023sample}. We include 17 such sources in our catalogue, with detailed descriptions provided in Appendix \ref{Appendix:extra_TDEs}.

In total our main catalogue consists of 134 TDEs (131 optical TDEs and three jetted TDEs), including all confirmed events up to 2024. As it was briefly mentioned in the Introduction, the creation of a catalogue of all the known optical TDEs so far can allow for statistical works, which were previously unable to be carried out due to small sample sizes. We note that this is not a complete sample, since many detected events remain unclassified and certain subtypes (e.g. events detected in IR) are not fully explored. 

\subsection{TDE candidates}\label{subsec:Honorable}

Alongside our main TDE sample, numerous transients have been classified as TDE candidates. These sources are included in a supplementary table, available on our GitHub page (see Sect. \ref{sec:data}). We note that this is not an exhaustive list of candidates \citep[e.g. we have not included the IR candidate TDEs from][]{Masterson2024IRcandidates} and the sources included are not used in any analysis throughout this work. The setup of this file is shown in Table \ref{tab:honorable}. The first four columns list the name and coordinates, while the fifth and sixth columns include remarks on the candidate status and reference studies, respectively. To compile the TDE candidates sample, along with our literature search, we utilised TDExplorer\footnote{\url{https://jminding.pythonanywhere.com/main/tdes-by-name}}, which is a catalogue of TDEs and candidates identified through natural language processing applied to the abstracts of papers.

\begin{table*}[htbp!]
	\caption{\label{tab:honorable}Setup of the TDE candidate sample table.}
    \centering
	\resizebox{\textwidth}{!}{
		\begin{tabular}{llcccl}
			\toprule
			AT Name    & Alternative Names               & RA           & DEC           & Comments                                   & Reference \\[3pt]
			\midrule
			-          & OGLE16aaa                       & 01:07:20.88  & -64:16:20.70  & TDE candidate in a weak AGN                & \citet{Wyrzykowski2017OGLE16aaa} \\[3pt]
			AT 2019eve & ZTF19aatylnl/Gaia19bti           & 11:28:49.650 & +15:40:22.30  & Persistent $H_\alpha$ line one year post peak. & \begin{tabular}[c]{@{}l@{}}\citet{vanVelzen2021sample},\\ \citet{Hammerstein2023ZTF1}\end{tabular} \\[3pt]
			AT 2021lwx & ZTF20abrbeie                    & 21:13:48.405 & +27:25:50.46  & Ultraluminous, long duration transient      & \citet{Subrayan2023AT2021lwx} \\[3pt]
			AT 2024kmq & ZTF24aapvieu                    & 12:02:37.273 & +35:23:35.22  & Luminous, fast, red transient                & \citet{Ho2025AT2024kmq} \\[3pt]
			-          & Swift J1112+82                  & 11:11:47.32  & -82:38:44.20  & Likely jetted TDE                          & \citet{Brown2015Sw1112+82} \\
			\bottomrule
		\end{tabular}
	}
    \tablefoot{We present an example for 5 TDE candidates of how the TDE candidate table is structured. Column 1: TNS name of the TDE candidate; Column 2: Alternative name; Columns 3--4: RA and DEC coordinates; Column 5: Relevant comments; Column 6: Reference.}
\end{table*}

\section{Data}\label{sec:data}

The catalogue is available online on a dedicated GitHub\footnote{\url{https://github.com/dlangis/TDECat}} page. On this page we have compiled all publicly available photometric and spectroscopic data for the TDEs in our sample. The full catalogue is also available via a local Python-based app. Below we summarise how and from where we collected the photometric and spectroscopic data included in the catalogue. We note that some artifacts could be present in the optical light curves (e.g. see light curve of AT 2019teq) included in the TDECat GitHub page.

\subsection{Optical and infrared photometry}\label{subsec:otpical data}

For compiling the optical and IR photometric data, we used the Black Hole Target Observation Manager (BHTOM\footnote{\url{https://bhtom.space}}), a web server designed to provide astronomers easy access to astronomical data and a network of telescopes. One of the key features of BHTOM allows the user to compile all the available, archived photometric data in an interactive plot and a .csv downloadable file. 
The catalogue includes data from several surveys, namely LINEAR, CRTS, ZTF, iPTF, SDSS, ASAS-SN, ATLAS, NEOWISE, Pan-STARRS and GSA.

We also manually searched for CRTS light curves of LSQ12dyw from the Catalina Surveys Data Release 2\footnote{\url{http://nesssi.cacr.caltech.edu/DataRelease/}} \citep{Drake2009CRTS}. 
Furthermore, for specific TDEs, we obtained archival photometric data from previously published studies:

\begin{itemize}
	\item PS1-10jh: Table S1 in the supplementary information of \citet{Gezari2012PS1-10jh}
	\item ASASSN-15oi: Table A1 of \citet{Holoien2016ASASSN15oi}.
	\item AT 2017eqx: Table 1 of \citet{Nicholl2019AT2017eqx}.
	\item iPTF16axa: Table A1 of \citet{Hung2017iPTF16axa}.
	\item iPTF15af: Table 2 of \citet{Blagorodnova2019iPTF15af}.
\end{itemize}

Since different surveys use different photometric methods, we converted all magnitudes and estimated flux densities in the AB system \citep{Oke1983AB} throughout this work. The monochromatic AB magnitude is defined as $m_{AB}\approx -2.5 log(\frac{f_\nu}{3631\ Jy})$, where $f_\nu$ is the spectral flux density and $3631\ \text{Jy}$ is the zero point. ZTF magnitudes are calibrated using a source with colour g - r = 0 in the AB photometric system. ATLAS, iPTF, Pan-STARRS and ASAS-SN $g-$band photometry is provided in the AB system. On the contrary, CRTS, ASAS-SN $V-$band, {\it Swift} and NEOWISE\footnote{IR zero points for conversion; \url{https://wise2.ipac.caltech.edu/docs/re   lease/allsky/expsup/sec4_4h.html}} photometry is given in the Vega system. For Vega-to-AB magnitude conversions, we used the values presented in Table 1 of \citet{Blanton2007vegaABconversions}. The Gaia survey uses an internal calibration different from both the AB and Vega systems.

\subsection{X-ray photometry}\label{subsec:X-ray phot}

To study the X-ray emission from TDE sources we used data from \textit{Swift}, \textit{Chandra}, and \textit{XMM-Newton} archives. For each TDE source, we analysed all observations available from these missions, and evaluated X-ray fluxes and spectra for each observation. In addition to \textit{Swift}-XRT, \textit{Chanra}-ACIS and \textit{XMM-Newton}-EPIC data, we searched for X-ray counterparts in the 13th data release of the fourth XMM-Newton serendipitous source catalogue \citep[4XMM-DR13,][]{2020AA...641A.136W} and eROSITA main catalogue \citep[eRASS1,][]{2024AA...682A..34M}. A detailed description of the X-ray data reduction is provided in Appendix \ref{Appendix:X-rays}. 

\subsection{Ultraviolet photometry}\label{sec:uvot}

To evaluate TDE variability in UV-optical bands, we used data from the Swift Ultraviolet and Optical Telescope \citep[UVOT;][]{2005SSRv..120...95R}, which provides photometry in three near-UV (UVW2, UVM2 and UVW1) and three optical (U, B, V) bands.

The UVOT data were downloaded from HEASARC\footnote{\href{https://heasarc.gsfc.nasa.gov/}{https://heasarc.gsfc.nasa.gov/}} and reduced using standard procedures\footnote{\href{http://www.swift.ac.uk/analysis/uvot/image.php}{http://www.swift.ac.uk/analysis/uvot/image.php}} (HEAsoft package v. 6.33.2). After checking the correct World Coordinate System alignment with USNO-B Catalogue \citep{2003AJ....125..984M}, we combined image extensions using \textsc{uvotimsum} and merged exposure maps.

Sources were detected using \textsc{uvotdetect}. As with XRT data (see Appendix \ref{sec:swift}), we assigned UVOT counterparts based on proximity to the TDE source coordinates: if a UVOT sources was detected within \(5\arcsec\), its coordinates were adopted. If no source was detected within \(5\arcsec\), we used the TDE source coordinates.

We performed source photometry using \textsc{uvotsource} with an aperture radius of \(5\arcsec\) for all filters. The background region was an annulus centred at the source with inner and outer radii of \(10\arcsec\) and \(15\arcsec\), respectively, removing emission from overlapping sources. All magnitudes were corrected for Galactic extinction using reddening estimates from \citet{2011ApJ...737..103S} and the extinction model from \citet{2007ApJ...663..320F}. We note that these data are not host subtracted.

\subsection{Spectroscopic data}

All optical spectra included in the catalogue were obtained from either TNS or the Weizmann Interactive Supernova Data Repository\footnote{\url{https://www.wiserep.org}} \citep[WISeREP; ][]{Yaron2012WISeREP}. Hence, the catalogue does not include optical spectra for TDEs not classified in TNS. For the 12 TDEs from \citet{Hammerstein2023ZTF1}, the classification spectra can be found in their Figs. 2 and 14-16. For AT 2018lni, AT 2019cho, AT 2019mha and AT 2019meg, the classification spectra can be found in \citet{vanVelzen2021sample}. 

Additionally, in the following we list references for (mainly optical) spectra of 13 TDEs that are not in the TNS-TDE sample or included in a previously published catalogue:

\begin{itemize}
	\item AT 2023vto: Fig. 4 in \citet{Kumar2024AT2023vto}.
	\item AT 2022agi: Figs. 2 (optical), 3 (UV) in \citet{sun2024recurringtidaldisruptionevents}.
	\item AT 2017gge: Fig. 2 (optical) in \citet{Wang2022AT2017gge}, Fig. 5 (near IR) in \citet{Onori2022AT2017gge}.
	\item AT 2017eqx: Figs. 1, 3 in \citet{Nicholl2019AT2017eqx}.
	\item LSQ12dyw.$^\dagger$\footnote{The $^\dagger$ symbol indicates the sources that have spectra available in WISeREP.}
	\item PTF09djl$^\dagger$: Figs. 4, 14 in \citet{Arcavi2014PTFTDEs}.
	\item PTF09ge$^\dagger$: Fig. 12, 14 in \citet{Arcavi2014PTFTDEs}.
	\item PS1-10jh$^\dagger$: Fig. 1 in \citet{Gezari2012PS1-10jh}.
	\item iPTF15af$^\dagger$: Figs. 3 (optical), 4 (UV) in \citet{Blagorodnova2019iPTF15af}.
	\item iPTF16axa$^\dagger$: Figs. 7, 8 in \citet{Hung2017iPTF16axa}.
	\item ASASSN-14li$^\dagger$: Fig. 3 in \citet{Holoien2016ASASSN14liOptXray}.
	\item ASASSN-14ae$^\dagger$: Figs. 4, 5 in \citet{Holoien2014ASASSN14ae}.
	\item ASASSN-15oi$^\dagger$: Fig. 3 in \citet{Holoien2016ASASSN15oi}.
\end{itemize}

\section{Exploring the catalogue}\label{sect:Explore}

After having constructed our main sample, we now explore its various subcategories and interesting objects. Specifically, we present the statistics and spectral classes of the TDEs in the catalogue, along with new results on TDEs with repeating flares, delayed IR emission and X-ray outbursts.

\subsection{Catalogue statistics}\label{subsect:statistics}
In this section we analyse the flaring characteristics of the TDE population, in terms of rise and decay times and the event durations. To achieve this, we applied the Bayesian Blocks \citep{Scargle1998BayesianBlocks,Scargle2013BayesianBlocks} algorithm, which partitions a one-dimensional flux time series into blocks, each modelled by a constant flux. To avoid overfitting, we include a user-specified penalty parameter for each additional block.

Given flux measurements $\{f_j\}$ and their uncertainties $\{\sigma_j\}$, we first compute cumulative weighted sums (with $w_j = 1/\sigma_j^2$) up to each index: $\sum w, \ \sum (w f), \ \sum (w f^2)$. These cumulative sums allow for efficient computation of sums over any sub-interval $[k, i)$ via $\text{Sum}([k,i)) = \text{Cumulative Sum}[i] - \text{Cumulative Sum}[k]$. Consequently, the maximum-likelihood estimate for the constant flux $\mu$ in a block spanning $[k, i)$ is $\mu = \frac{\sum_{j=k}^{i-1} \bigl(w_j f_j\bigr)}{\sum_{j=k}^{i-1} w_j}$.

Assuming Gaussian errors, the log-likelihood of modelling data in the interval $[k,i)$ with a single constant value $\mu$ is
\begin{center}
	\begin{equation}
		\log\bigl(\mathcal{L}\bigr) \;=\; - \tfrac{1}{2} \sum_{j=k}^{i-1} \frac{\bigl(f_j - \mu\bigr)^2}{\sigma_j^2}.
	\end{equation}
\end{center}
By expanding the summation in the exponent, one obtains $- \tfrac{1}{2} \Bigl[\sum_{j=k}^{i-1} w_j f_j^2 \;-\; 2\,\mu\sum_{j=k}^{i-1} w_j f_j \;+\; \Bigl(\sum_{j=k}^{i-1} w_j\Bigr)\,\mu^2 \Bigr].$

Moreover, the algorithm uses dynamic programming to determine the optimal segmentation. For each index $i$ (ranging from 1 to the total number of data points $n$), all possible previous boundaries $k$ (from $0$ to $i-1$) are considered. We computed

\begin{center}
	\begin{equation*}
		\text{candidate\_score} = \text{best}[k] \;+\; \log\bigl(\mathcal{L}_{k \to i}\bigr) \;-\; \text{penalty},
	\end{equation*}
\end{center}

\noindent where 

\begin{itemize}
	\item $\text{best}[k]$ is the optimal (maximum) log-likelihood score obtained for the segmentation 
	of the interval $[0,k)$,
	\item $\log(\mathcal{L}_{k \to i})$ is the log-likelihood for modelling the new block 
	$[k,i)$ with constant flux, and
	\item $\text{penalty}$ is a user-specified constant subtracted each time a new block is added. We chose a default value of ten for the penalty to avoid overfitting or underfitting (i.e. small or large penalty values respectively). 
\end{itemize}
The optimal segmentation was obtained by choosing the $k$ that maximises $\text{candidate\_score}$ followed by a backtracking step (from $i=n$ to $0$) to recover the optimal boundaries.

Applying the Bayesian Blocks algorithm to the flux light curves yields blocks within which the flux is assumed to be constant. For each block, we record its mean flux value. The peak of the flare is then identified as the block with the highest mean flux. We primarily use ZTF(zg) light curves (76/103). In case ZTF data are not available or the ZTF(zg) light curve is poorly sampled, we used the light curve with the best sampling. This resulted in 11 ZTF(zr) light curves, six ATLAS light curves, five ASAS-SN light curves, one GSA(G) light curve, one PS1 light curve, one CRTS(CL) light curve, and two PTF light curves.

\begin{figure}[t!]
	\centering
	\includegraphics[width=\hsize]{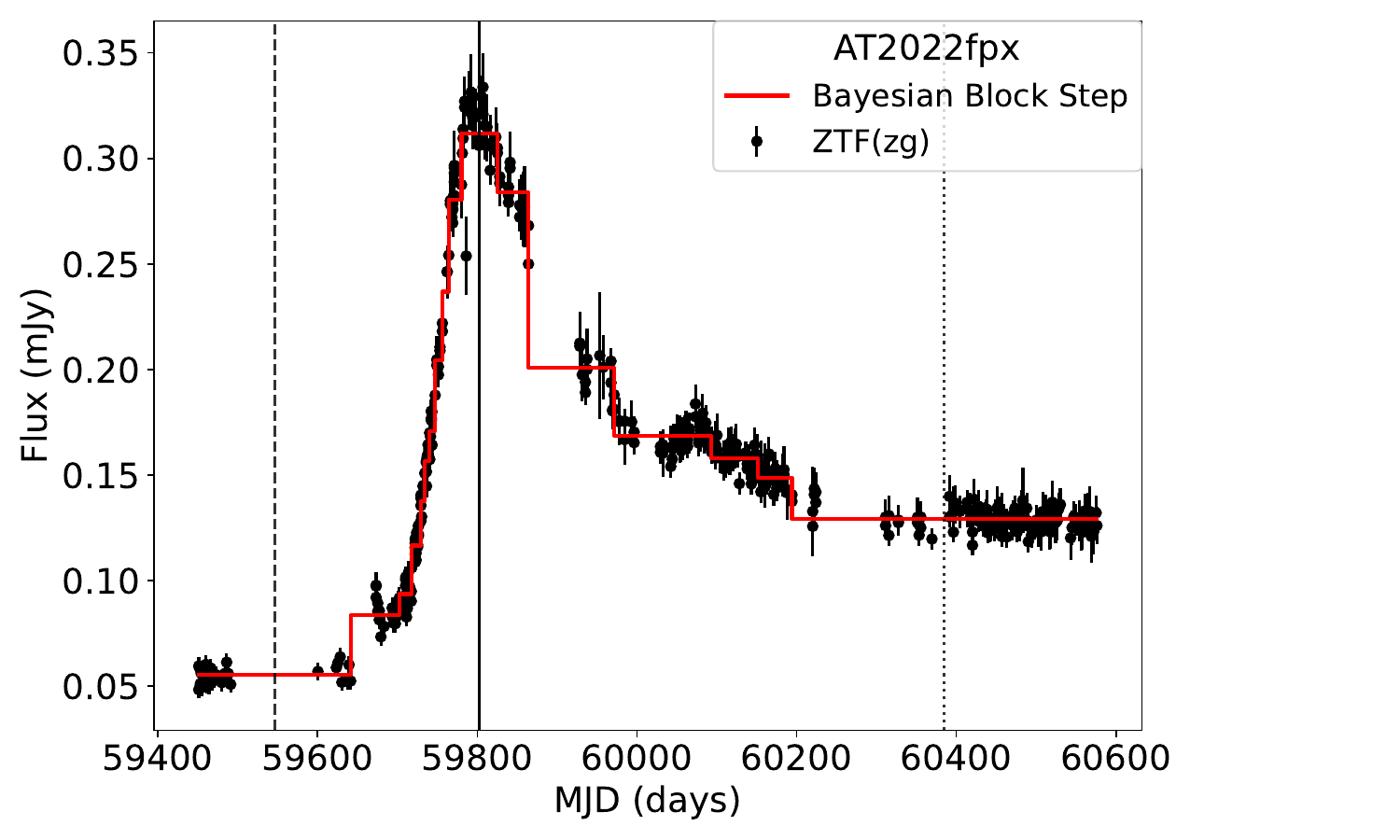}
	\caption{Example of the implementation of the Bayesian Blocks algorithm to the ZTF light curve of AT 2022fpx. The red solid line shows the optimal blocks and the dashed, dotted and solid black lines represent the rise, decay, and peak times, respectively.}
	\label{fig:BBex}
\end{figure}

To estimate the rise and decay times of the flare, we apply a $2.5\%$ amplitude-based threshold (amplitude = peak flux - minimum flux). Starting from the peak block, we move to lower flux levels (to lower or higher block indices) until encountering a block with a mean flux below this threshold. The first sub-threshold block on the left marks the start of the rise, while the first sub-threshold segment on the right marks the end of the decay. From these boundaries, we calculate the rise time, $t_{\mathrm{rise}}$, and decay time, $t_{\mathrm{decay}}$, and determine their ratio $t_{\mathrm{rise}}/t_{\mathrm{decay}}$. We note that in cases where the noise of the data is larger than the $2.5\%$ amplitude-based threshold, the flare is not constrained to when it drops to $2.5\%$, but where it reaches the observational noise.

An example of this process is displayed in Fig. \ref{fig:BBex}. For this example, the rise time is 256$\pm$98 days, the decay time 582$\pm$192 days and the peak time 59803$\pm$22 modified Julian date. The duration of the flare, calculated as the sum of the rise and decay times, is roughly 838 days. 

\begin{figure}[t]
	\centering
	\includegraphics[width=\hsize]{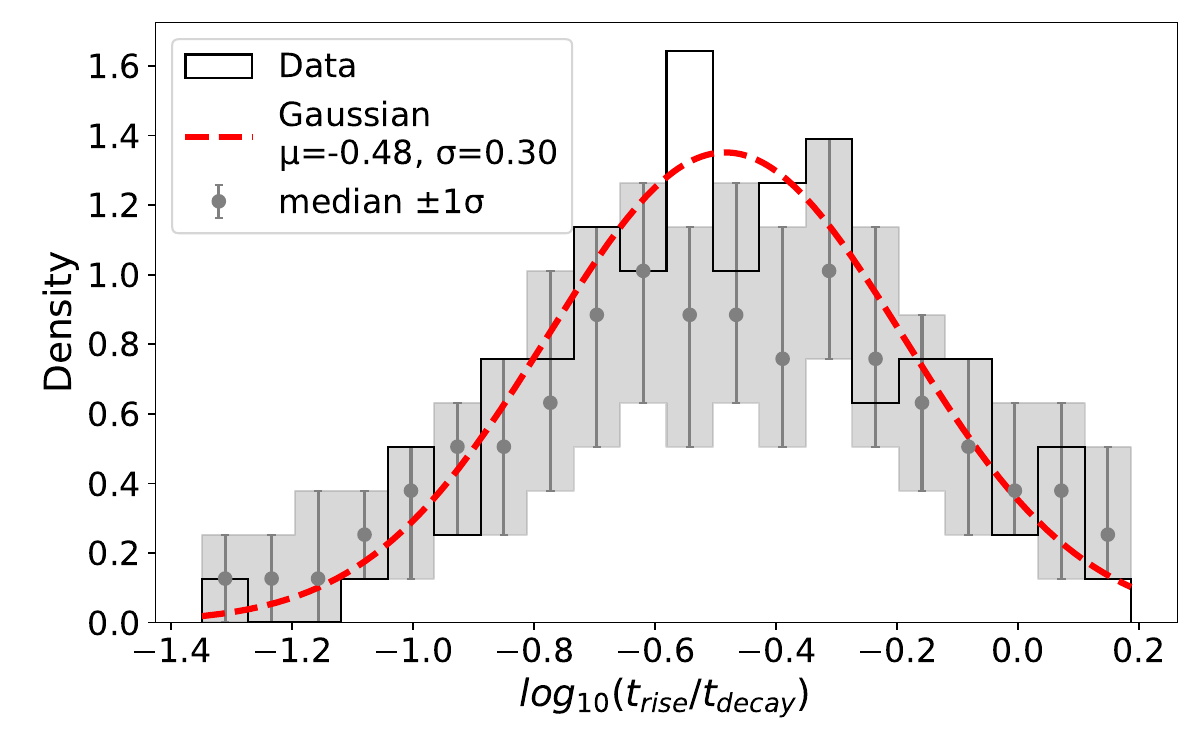}
	\caption{Distribution of the $t_{rise}/t_{decay}$ ratio for 103 TDE flares in our catalogue (black step histogram). The dashed red line indicates a Gaussian distribution with $\mu=-0.48$ and $\sigma=0.30$ overplotted on the data. The grey data points and boxes represent the median of the mock distributions for each bin and corresponding uncertainty respectively.}
	\label{fig:ratio_distribution}
\end{figure}

We applied this method to 103 TDE flares in our sample, excluding most of the sources detected in 2024 and some older ones where rise or decay times could not be reliably estimated due to large gaps in the light curves. We also included both flares for AT 2022dbl and AT 2020vdq (see Sect. \ref{subsect:Repeating}). Figure \ref{fig:ratio_distribution} shows the distribution of the common logarithm (i.e. logarithm with base 10) of the ratio $t_{\mathrm{rise}}/t_{\mathrm{decay}}$ in our sample. A Kolmogorov-Smirnov (KS) test of the data against a normal distribution with $\mu=-0.48$ and $\sigma=0.30$ (the mean and standard deviation of the data respectively) yields a p-value of 0.99 (KS statistic = 0.04). This means that the $t_{\mathrm{rise}}/t_\mathrm{decay}$ ratio distribution is consistent with a log-normal distribution.

\begin{figure*}[htbp!]
	\centering
	\includegraphics[width=\hsize]{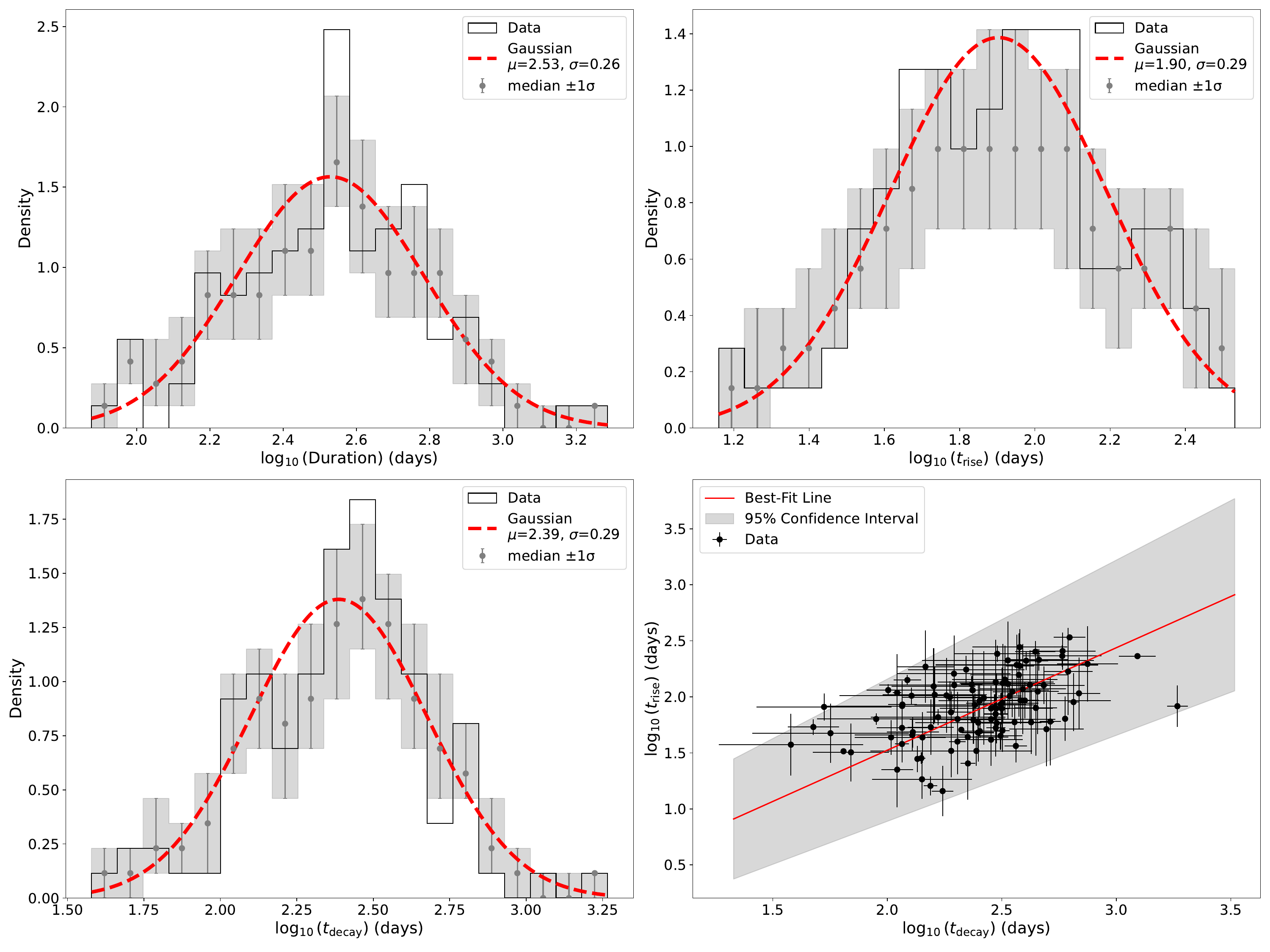}
	\caption{Distributions of the durations, rise times, and decay times of the flares (top panels and bottom-left panel, respectively), as well as the scatter plot of $t_{rise}$ versus $t_{decay}$ (bottom-right panel). The red dashed lines indicate Gaussian distributions overplotted on the data. The grey data points and boxes represent the median of the generated distributions for each bin and corresponding uncertainty, respectively. In the bottom-right panel the solid red line indicates the best-fit ODR line, while the grey contour represents the 95\% confidence interval.}
	\label{fig:all_distributions}
\end{figure*}

Furthermore, we estimated the statistical errors of the histogram in Fig. \ref{fig:ratio_distribution} (and Fig. \ref{fig:all_distributions}). First, we set a fixed width for the bins of the histogram. Then we computed $10^5$ realisations of mock histograms by generating new values from a normal distribution of $\mu$ equal to each observed value (black step histogram data) and $\sigma$, its corresponding error. This resulted in $10^5$ mock histograms, with different bin heights. We computed the median of all the realisations for each bin along with its uncertainty (the distance from the median to the 16th percentile for the lower error and the distance from the 84th percentile to the median for the upper error), plotted as the grey points and grey boxes in Fig. \ref{fig:ratio_distribution} (and Fig. \ref{fig:all_distributions}).

Figure \ref{fig:all_distributions} displays the duration, rise and decay times of the flares. We utilised KS tests against Gaussian distributions with means\footnote{ When back-transforming from log space to linear space, smearing retransformation needs to be applied \citep{Duan1983SmearingFactor}.} of $\mu=$400$\pm$30 days, $\mu=$100$\pm$6 days and $\mu=$300$\pm$20 days for the duration, $t_{rise}$ and $t_{decay}$ (the error is the standard error of the mean) and standard deviations of $\sigma=$260 days, $\sigma=$74 days and $\sigma=$229 days respectively (which correspond to the mean and standard deviation of the data). The p-values are 0.87, 0.91 and 0.58 (KS statistic of 0.057, 0.054 and 0.075) for the durations, $t_{rise}$ and $t_{decay}$ respectively, consistent with log-normal distributions. We repeat the same process for the statistical errors of the histograms as presented in Fig. \ref{fig:ratio_distribution} for Fig. \ref{fig:all_distributions}.

The bottom-right panel of Fig. \ref{fig:all_distributions} shows a scatter plot of $log_{10}(t_{rise})\ \text{versus}\ log_{10}(t_{decay})$, with error bars calculated as $\sqrt{\sigma_{rise\ or \ decay}^2+\sigma_{peak}^2}$, where $\sigma_{rise\ or\ decay}$ is half the width of the rise or decay blocks and $\sigma_{peak}$ is half the width of the peak block. Given errors in both the independent [$log_{10}(t_{decay})$] and dependent [$log_{10}(t_{rise})$] variables, we employed the orthogonal distance regression (ODR) to fit a best-fit line while accounting for both uncertainties. The best-fit ODR line is shown as the red solid line in the bottom-right panel of Fig. \ref{fig:all_distributions}, while the grey contour indicates the 95\% confidence interval. It is evident that there is large scatter around the best-fit line, $log_{10}(t_{rise}) = \alpha log_{10}(t_{decay})+\beta$, where $\alpha = 0.915 \pm 0.075$ and $\beta = -0.31 \pm 0.17$, with a 95\% confidence interval of 0.768 to 1.062 and -0.65 to -0.04 for $\alpha$ and $\beta$ respectively. We tested whether the correlation between $log_{10}(t_{rise})$ and $log_{10}(t_{decay})$ is real using a t-test, the Spearman correlation coefficient ($\rho=0.48$, with a corresponding p-value $\approx3.6\times10^{-7}$), and bootstrapping. In all our tests we could not reject the null hypothesis of a strong, positive correlation between the rise and the decay times.

\begin{figure}[htbp!]
\centering
\includegraphics[width=\hsize]{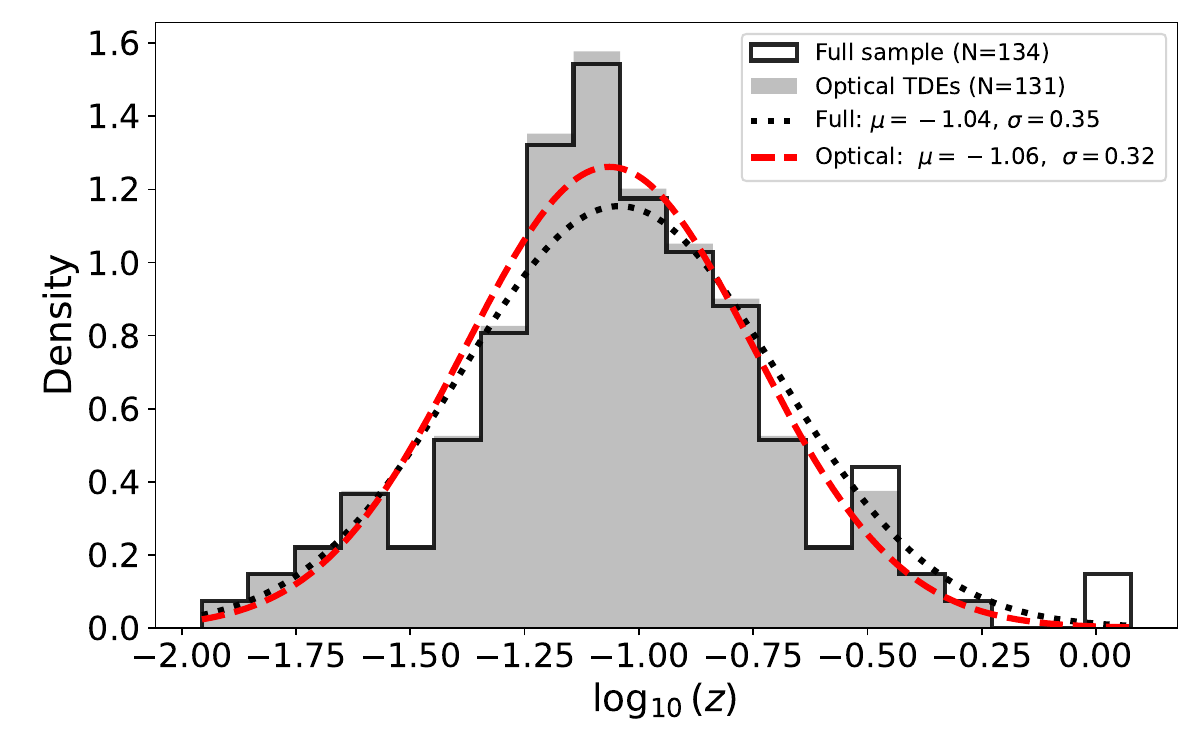}
\caption{Distribution of redshift for the TDEs in our sample (black step histogram) as well as the optical TDEs (grey step-filled histogram). The dotted black line indicates a normal distribution overplotted on the full sample, while the dashed red line represents a normal distribution overplotted on the optical TDEs.}
\label{fig:redshift_distribution}
\end{figure}

Figure \ref{fig:redshift_distribution} shows the redshift distributions for the TDEs in the main sample (black step histogram), as well as the optical TDEs (grey step-filled histogram). Both distributions also follow normal distributions in log-space with $\mu=0.13,\ \sigma=0.16$ (full sample) equal to the mean and the standard deviation of the data respectively (KS statistic equal to 0.0579 and $p=0.74$) and $\mu=0.11,\ \sigma=0.09$ for the optical TDEs (KS statistic equal to 0.0532 and $p=0.83$). We discuss further in Sect. \ref{subsect:stat discuss} about potential selection effects that could cause these distribution profiles, along with their use for generating mock data.

Additionally, we used the Anderson-Darling (AD) normality test to further examine the validity of the log-normal behaviour of the timescale and redshift distributions. In all cases, we could not reject the null hypothesis that the distributions are normal (in log-space).

\subsection{Spectral classification}\label{subsect:Spectral Type}

In this section we investigate the different spectral classes of the TDEs in our sample. For consistency, we adopted the three spectral classes defined in \citet{vanVelzen2021sample} as well as the TDE-featureless class introduced by \citet{Hammerstein2023ZTF1}:

\begin{itemize}
\item TDE-H: In this case, the spectrum exhibits distinct and broad H$\alpha$ and H$\beta$ emission lines.
\item TDE-H+He: Here, the spectrum shows both broad H$\alpha$ and H$\beta$ emission lines and broad He II emission features.
\item TDE-He: For this case, the only distinct broad feature in the spectrum appears near the He II emission line with no detectable Balmer emission lines.
\item TDE-featureless: Here, the spectrum primarily displays host absorption lines with no distinct emission features characteristic of the other three classes.
\end{itemize}

\begin{figure}[htbp!]
\centering
\includegraphics[width=\hsize]{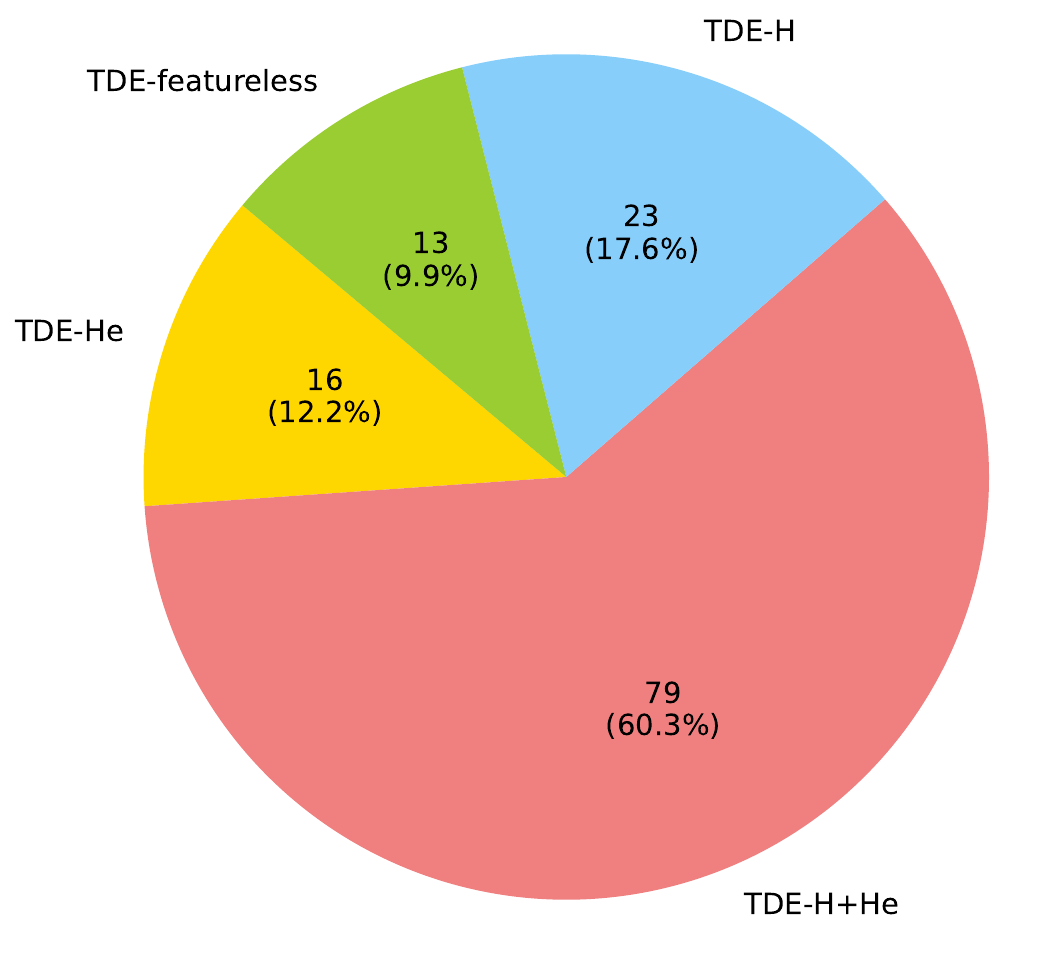}
\caption{Percentage of each TDE spectral class in the catalogue.}
\label{fig:pie chart spec}
\end{figure}

We note that for most objects in our sample only the classification spectrum is available, typically from the time of the flare. This makes the spectral classification uncertain, as the spectral properties of some TDEs evolve over time \citep[e.g. ][]{Charalampopoulos2022LSQ12dyw}. Several previous studies have already classified the TDEs in their samples, including \citet{vanVelzen2021sample}, \citet{Hammerstein2023ZTF1}, and \citet{Yao2023sample}. In some cases, the resulting classifications are mixed. For example, AT 2019mha, AT 2019bhf, and AT 2018hyz appear in two or even all of the above studies, but have different classifications. Additionally, \citet{Charalampopoulos2022LSQ12dyw} provided spectral classifications for the TDEs in their sample (see their Table 4), some of which contradict the classifications assigned by \citet{vanVelzen2021sample} for overlapping sources (namely ASASSN-15oi and PTF09ge). Furthermore, \citet{Charalampopoulos2022LSQ12dyw} identified three TDEs: AT 2018hyz, AT 2017eqx and ASASSN-14ae, that show spectral evolution over time (see their Table 4). Throughout this analysis, we assign events to the TDE-H+He class if they have been classified as such in at least one observation epoch. Additionally, six more transients listed in TNS have already been assigned a spectral class (three TDE-H+He and three TDE-He). Moreover, spectral classifications or information on the broadness of the emission lines are available in TNS AstroNotes or classification reports of several TDEs. Finally, since the classification spectra are publicly available for almost all sources in the catalogue, we classify the remaining TDEs in this work.

The results of this analysis are presented in Fig. \ref{fig:pie chart spec}. To summarise, 17.6\% (23) of the TDEs belong to the TDE-H class, 12.2\% (16) are classified as TDE-He, 60.3\% (79) fall into the TDE-H+He class and 9.9\% (13) are TDE-featureless. We note that even when considering TDEs with multiple classifications based on spectra from different epochs, the overall distribution of spectral classes remains largely unchanged. If we consider the alternative classifications (PTF09ge$\rightarrow$TDE-He, ASASSN-14ae$\rightarrow$TDE-H, ASASSN-15oi$\rightarrow$TDE-He, AT 2017eqx$\rightarrow$TDE-He, AT 2018hyz$\rightarrow$TDE-H, AT 2019bhf$\rightarrow$TDE-H, AT 2019mha$\rightarrow$TDE-H), the aforementioned percentages change to 20.6\% (27), 13.7\% (18), 55.7\% (73) and 9.9\% (13) for the TDE-H, TDE-He, TDE-H+He and TDE-featureless classes. The spectral classification of each TDE, along with the corresponding references, can be found on the catalogue GitHub page. We discuss the implications of our results in Sect. \ref{subsec:spect_class_disc}

\subsection{Repeating flares}\label{subsect:Repeating}

\begin{figure*}[htbp!]
\centering

\begin{subfigure}{0.48\textwidth}
	\centering
	\includegraphics[width=\linewidth]{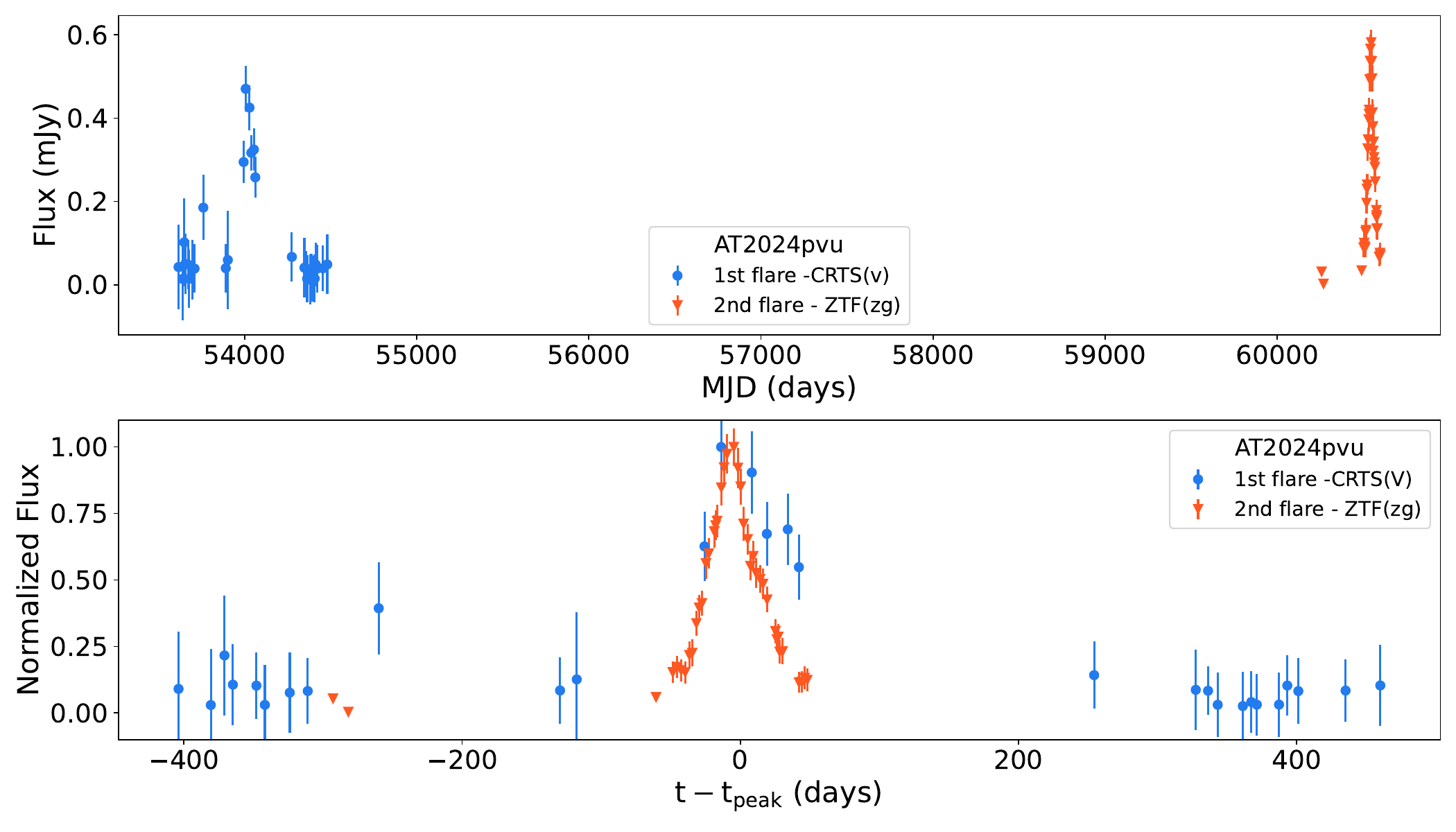}
	\caption{}
	\label{fig:subfigAT2024pvu}
\end{subfigure}
\hfill
\begin{subfigure}{0.48\textwidth}
	\centering
	\includegraphics[width=\linewidth]{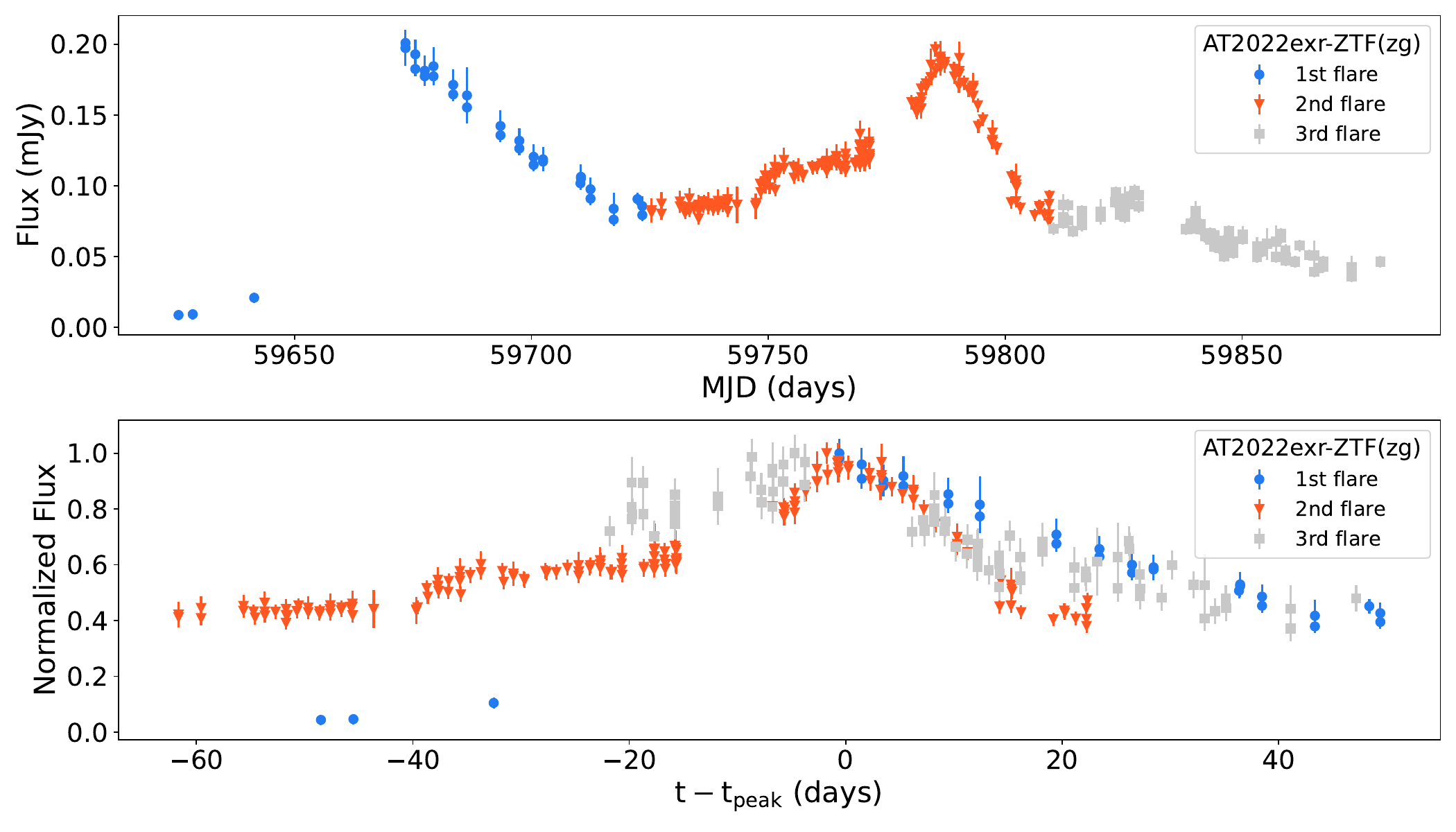}
	\caption{}
	\label{fig:subfigAT2022exr}
\end{subfigure}

\begin{subfigure}{0.48\textwidth}
	\centering
	\includegraphics[width=\linewidth]{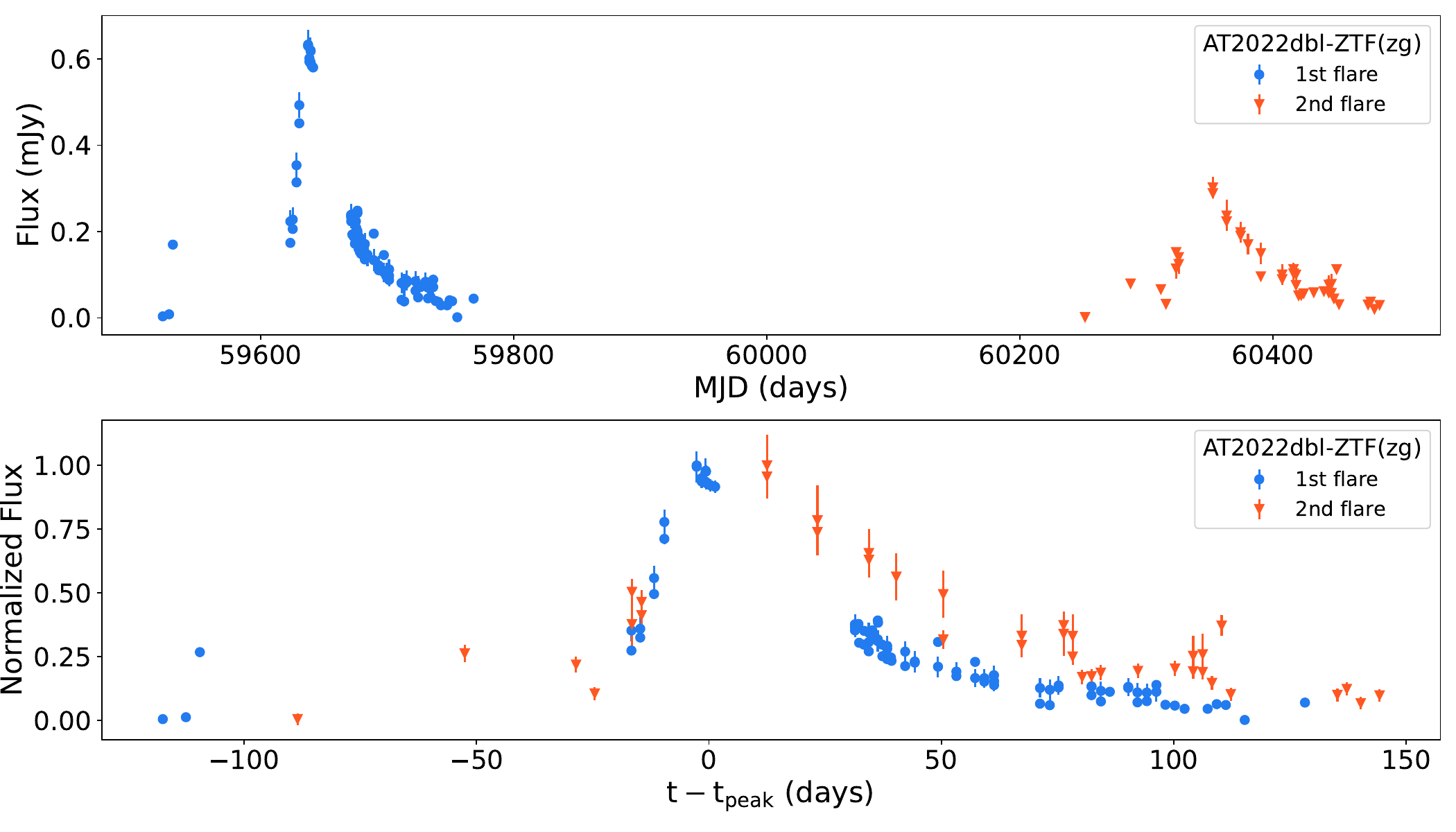}
	\caption{}
	\label{fig:subfigAT2022dbl}
\end{subfigure}
\hfill
\begin{subfigure}{0.48\textwidth}
	\centering
	\includegraphics[width=\linewidth]{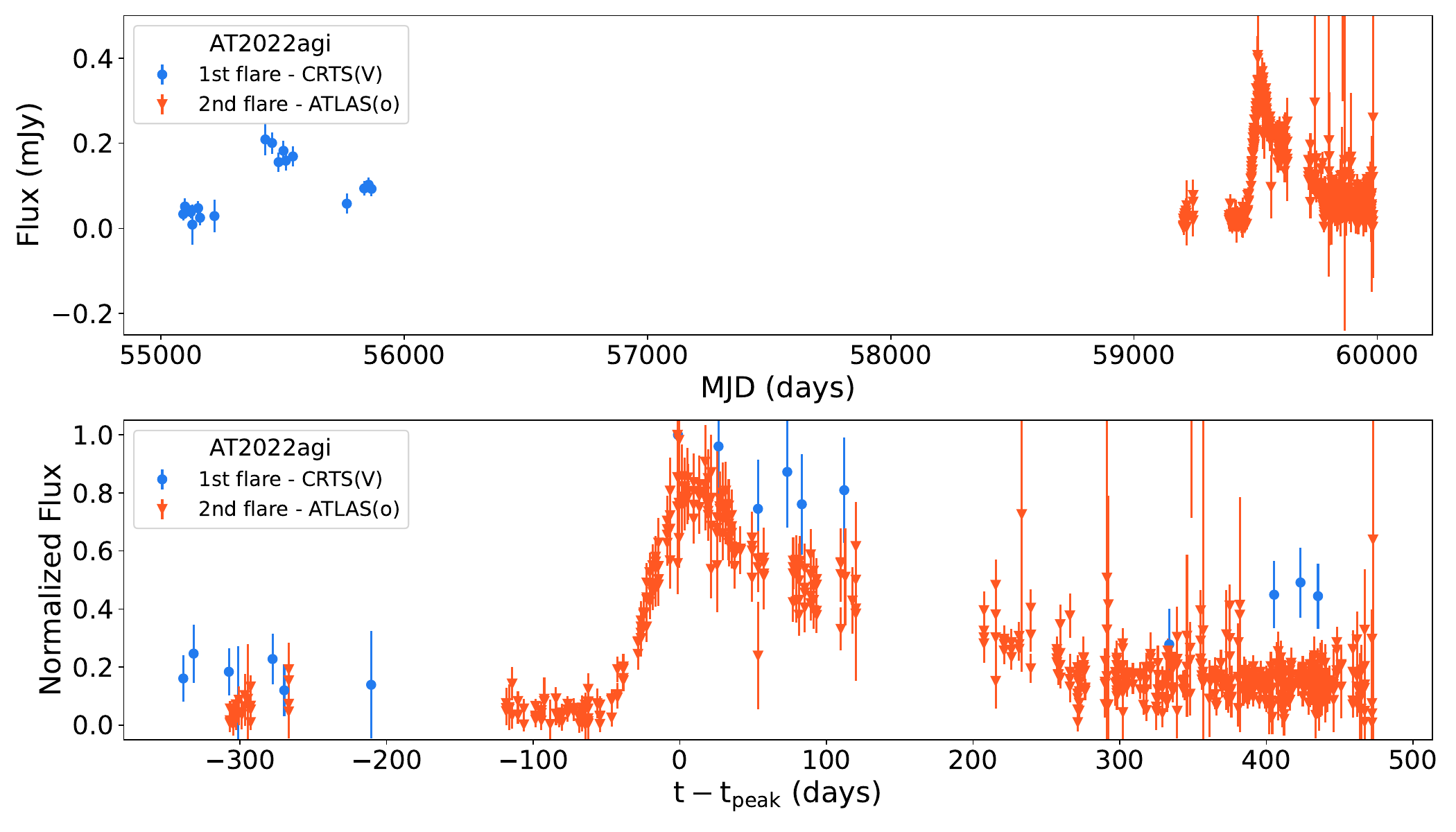}
	\caption{}
	\label{fig:subfigAT2022agi}
\end{subfigure}

\begin{subfigure}{0.48\textwidth}
	\centering
	\includegraphics[width=\linewidth]{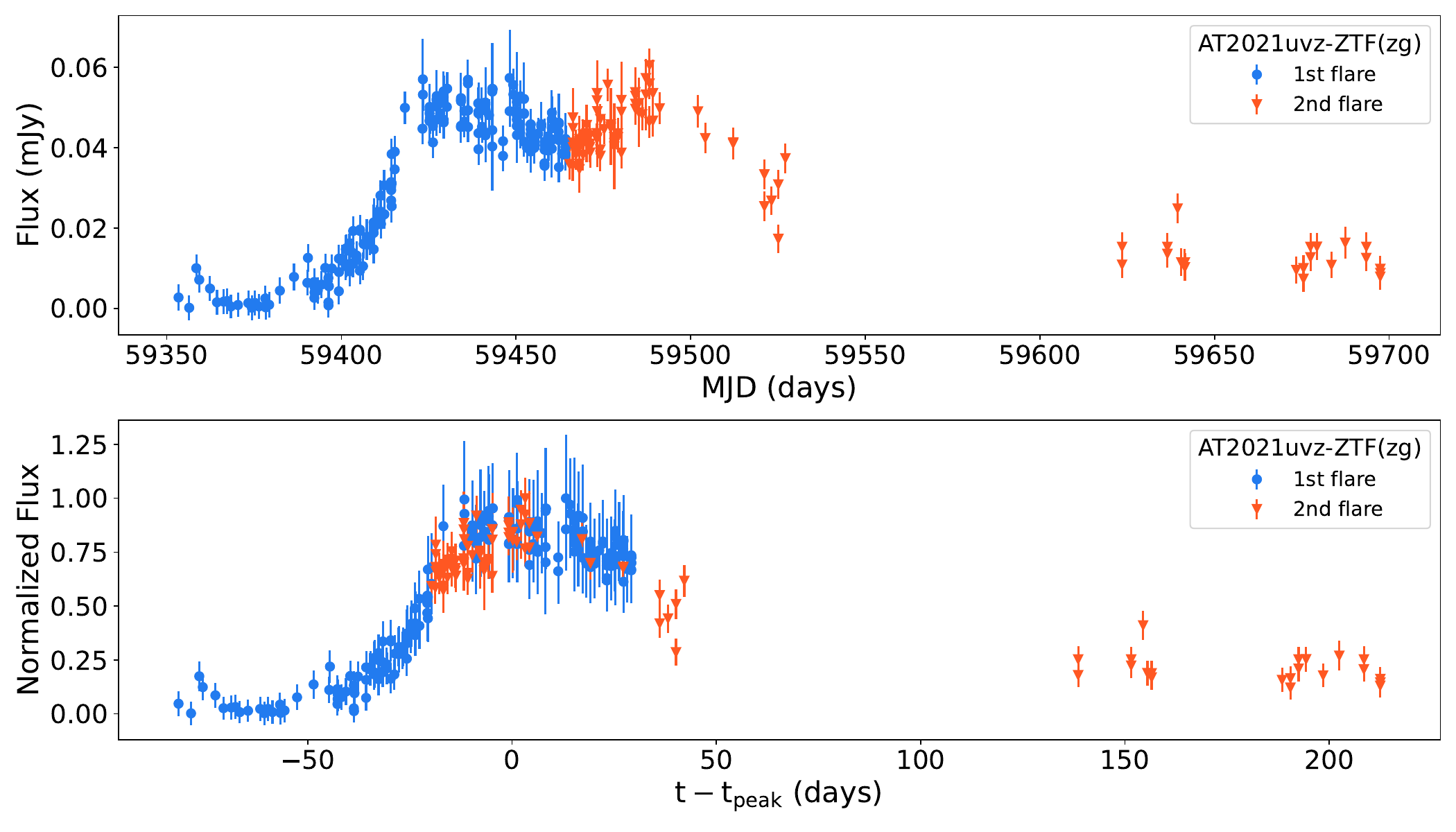}
	\caption{}
	\label{fig:subfigAT2021uvz}
\end{subfigure}
\hfill
\begin{subfigure}{0.48\textwidth}
	\centering
	\includegraphics[width=\linewidth]{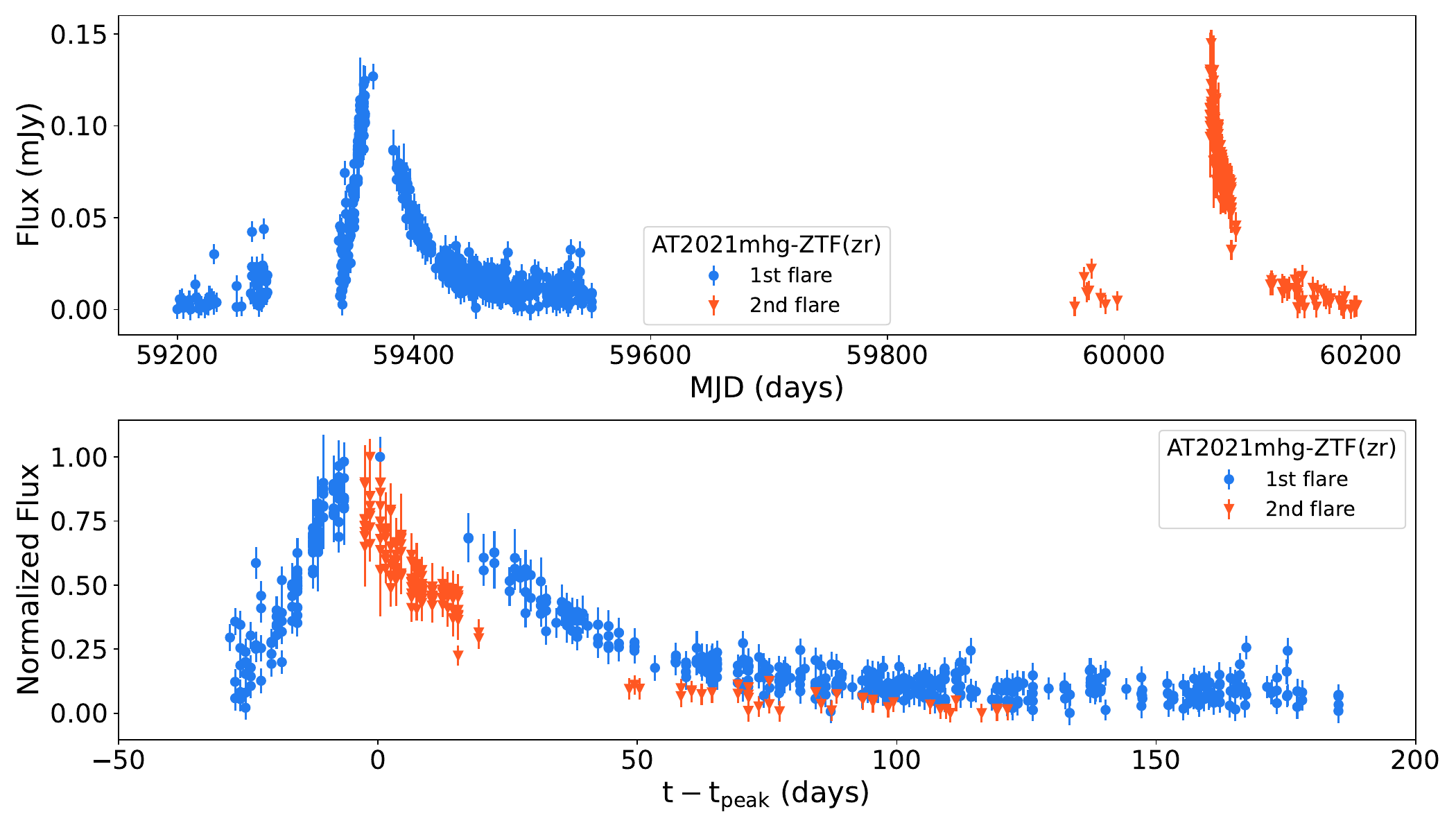}
	\caption{}
	\label{fig:subfigAT2021mhg}
\end{subfigure}

\caption{Optical light curves of the repeating flare TDEs shown in the different panel pairs (a-l). The initial flares are plotted using filled blue circles, with the following flares plotted using filled orange triangles and filled grey squares. In all panel pairs the upper panel shows the optical light curves of the flare regions from different surveys, while the bottom panel displays the two flares, shifted with respect to the flare peak in time and flux.}
\label{fig:alll repeating}
\end{figure*}

\begin{figure*}[htbp!]
\ContinuedFloat
\centering

\begin{subfigure}{0.48\textwidth}
	\centering
	\includegraphics[width=\linewidth]{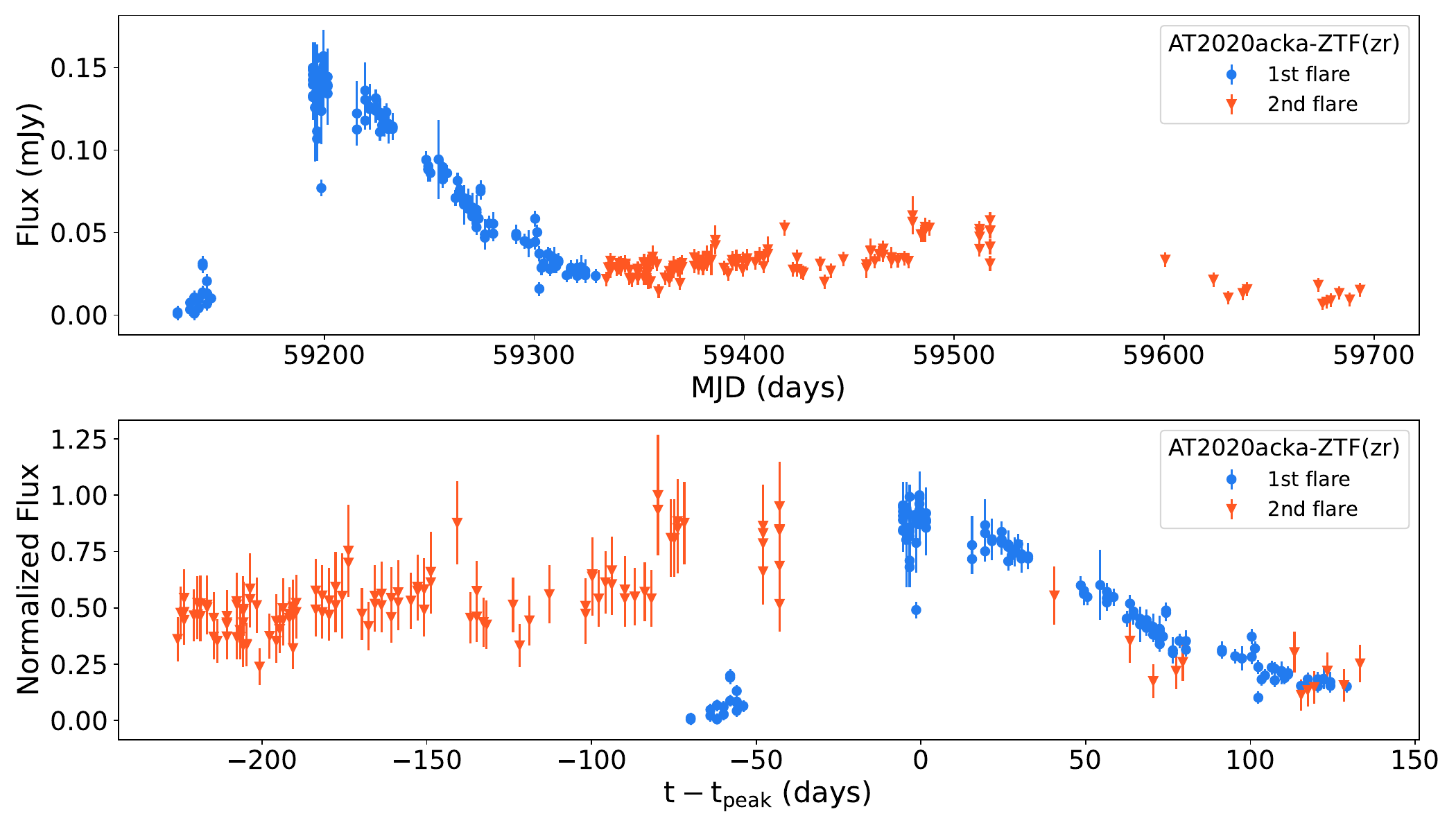}
	\caption{}
	\label{fig:subfigAT2020acka}
\end{subfigure}
\hfill
\begin{subfigure}{0.48\textwidth}
	\centering
	\includegraphics[width=\linewidth]{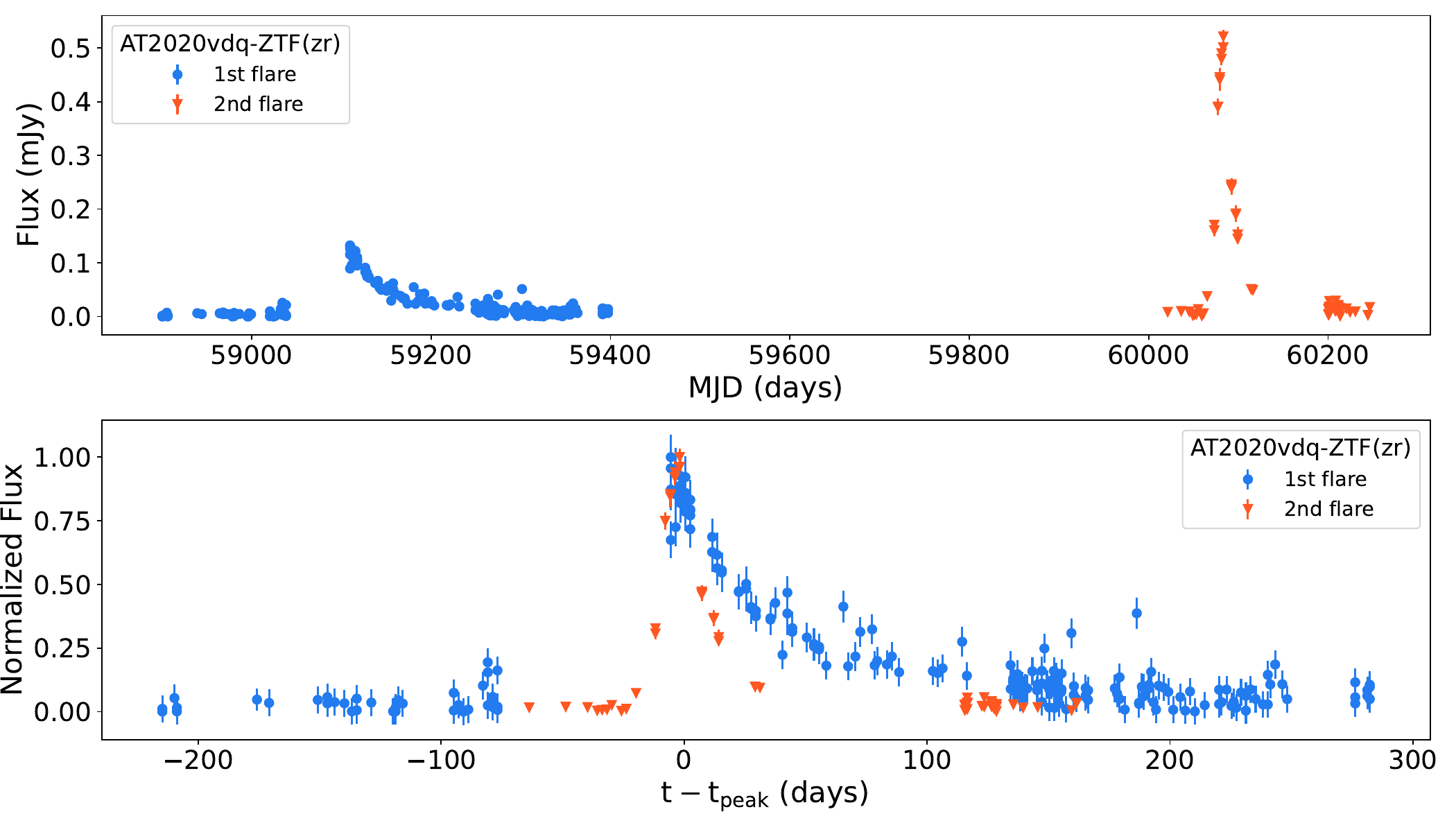}
	\caption{}
	\label{fig:subfigAT2020vdq}
\end{subfigure}

\begin{subfigure}{0.48\textwidth}
	\centering
	\includegraphics[width=\linewidth]{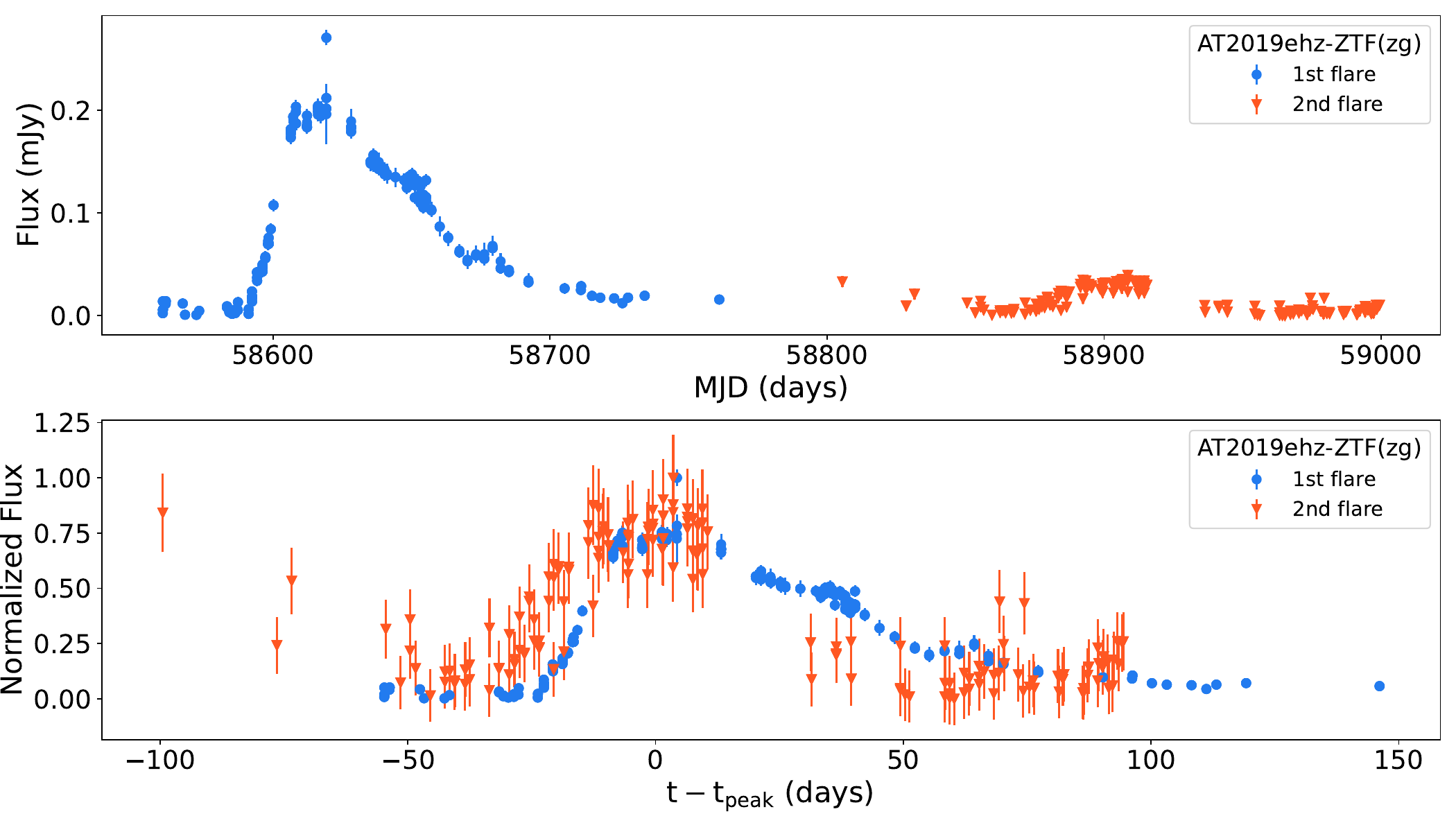}
	\caption{}
	\label{fig:subfigAT2019ehz_g}
\end{subfigure}
\hfill
\begin{subfigure}{0.48\textwidth}
	\centering
	\includegraphics[width=\linewidth]{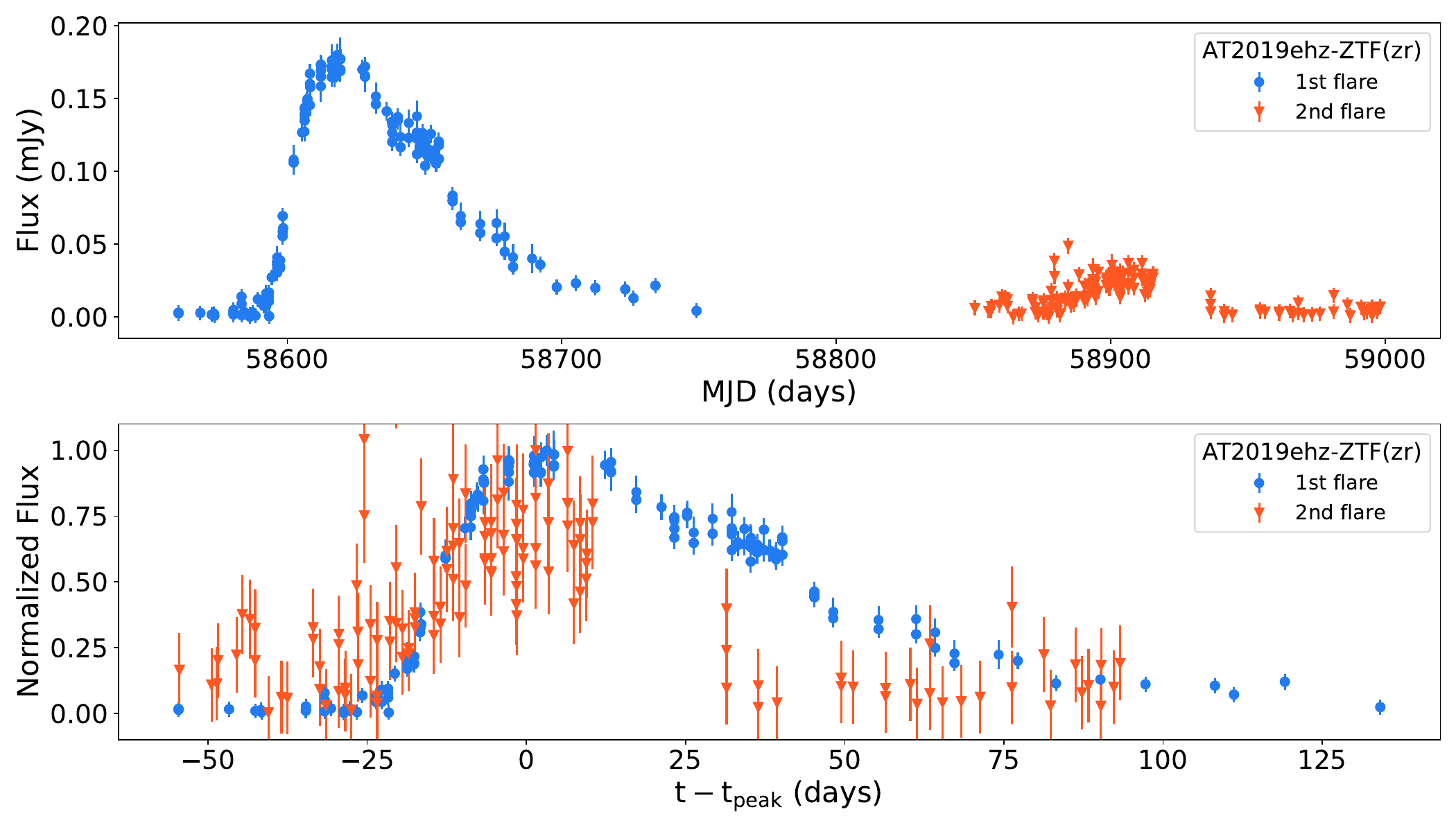}
	\caption{}
	\label{fig:subfigAT2019ehz_r}
\end{subfigure}
\caption{Continued.}
\end{figure*}

A few TDEs in our catalogue exhibit repeating flares. These flares may correspond to separate disruption events, where different stars are disrupted each time, or they may result from the partial disruption of the same star. In some cases, the re-brightening could be caused by emission from an accretion disc, towards the later stages of the flare. Additionally, AT 2021mhg represents a unique case believed to be a TDE followed by a supernova \citep[SN;][]{Somalwar2023AT2020vdq_however_AT2021mhg_appendix}. Below, we present our sample of repeating-flare TDEs and our analysis of three previously unreported cases: AT 2024pvu, AT 2022exr, and AT 2021uvz. To the best of our knowledge, these TDEs have not yet been reported as exhibiting a repeating flare.

In total we identified 11 TDEs with multiple flares. The corresponding light curves highlighting the repeating outbursts are shown in Fig. \ref{fig:alll repeating}. AT 2018fyk is plotted separately in Fig. \ref{fig:allAT2018fyk}, since it is the only TDE with Gaia photometry, which cannot be converted to flux. We select the best sampled light curves from the available surveys to maximise the visibility of the flares, plotting only the relevant flare regions.

We approximate the host galaxy emission as the average flux measured prior to the flare. To test the validity of this assumption, we split the light curves into bins corresponding to different ZTF observational seasons for each TDE, compute their average flux and standard deviation and compare the averaged values to the pre-flare flux using a $\chi^2$ test.

In all cases, the reduced $\chi^2$ values and corresponding p-values are consistent with the flux being constant, indicating that the pre-flare light curves can be approximated by the average flux. This average flux is then subtracted from the entire light curve to remove the host galaxy's contribution, retaining only the positive flux variations associated with the transient event. The host-subtracted light curves are displayed in Fig.~\ref{fig:alll repeating}.

The bottom panels of Fig.~\ref{fig:alll repeating} illustrate the multiple flares of each TDE, overlaid for direct comparison. We normalise the flux to the corresponding flare peak for comparison. We observe that the TDEs in the repeating flare sub-sample display flares that are similar in shape with the confirmed TDE flares. Most notably, we find AT 2024pvu to be the record holder for the longest time separation between two flares ($\sim 17.9$ yr), as well as that AT 2022exr and AT 2021uvz display a double-peak morphology.

For each of the repeating flare TDEs, we examine the possible mechanisms that could explain their multiple flares. We investigate the repeating partial TDE scenario (pTDE), the disruption of a stellar binary (double TDE) and the disruption of separate, unrelated stars. We also discuss alternative mechanisms for specific cases. A summary of this analysis is presented in Table \ref{tab:repeating}, where the first three columns show the name, the time separation between flares ($\Delta t$; taken from peak to peak) and the number of flares ($N_{flare}$) respectively, while the final four columns display the different scenarios we considered for the repeating nature of the events ($\checkmark \rightarrow $likely,$\ ?\rightarrow$ could explain the event but not very probable,$\ X\rightarrow$ not likely). The relevant references are listed in the final column.

First we explored the separate event scenario for the 11 sources in the repeating flare sample. In order to achieve this, we follow the analysis performed in \citet{sun2024recurringtidaldisruptionevents} for AT 2022agi (IRAS F01004-2237; see their Sect. 5.4). The probability of a separate event occuring in the same host after a time interval of $\Delta t$ can be computed by $p=r_{TDE}\times \Delta t$, where $r_{TDE}=3.2^{+0.8}_{-0.6}\times 10^{-5}$ yr$^{-1}$  galaxy$^{-1}$, which is the incident optical TDE rate from \citet{Yao2023sample}. For each TDE we consider the $\Delta t$ displayed in the second column of Table \ref{tab:repeating}, whereas the probability for each host galaxy to have another separate event in $\Delta t$ is given in the parenthesis of the sixth column. We observed that the smallest value is $4.8\times 10^{-6}$, while the largest is $5.7\times10^{-4}$. For $p$ to exceed $>0.05$, the TDE rate would need to be roughly 87, 152 and 108 times higher for AT 2024pvu, AT 2022agi and AT 2019azh respectively, while for the rest it would need to be 422 times higher or more, reaching $\sim 10370$ times for AT 2021uvz. For the first three we list, the required rate could be theoretically possible given the enhanced rates in post starburst galaxies, which can have boosted rates of up to $\sim200$ times on average \citep[e.g.][]{French2016ApJPostStarburst,Law-Smith2017ApJPostStarburst,Graur2018ApJPostStarburst}. However, for the latter cases, we can confidently exclude the scenario that the two flares were caused by separate events, since their rates would need to be enhanced by more than 400 times. We note that the reported p-values do not account for the look-elsewhere effect and could be substantially higher (even by a factor of 100 depending on the event). The correct treatment of the look-elsewhere effect would require dedicated simulations which are beyond the scope of this work.

In order to study whether the multiple flares could be explained by the partial disruption of the same star, we utilised a toy model to calculate the eccentricity. If we assume a solar-like star, a SMBH with a mass of $10^6~M_\odot$ and an orbital period of $\Delta t$, we retrieve orbits between the range $1-e_{max}\approx0.0007$ and $1-e_{min}\approx0.02$. Such orbits would be highly unstable, making the pTDE scenario unlikely. However, such highly eccentric orbits could be explained by the Hills Mechanism \citep{Hills1988NaturHillsMechanism}, where the SMBH tidally breaks a stellar binary: one star is ejected with hyper-velocity while the other is captured on an extreme eccentric bound orbit. We use another toy model to qualitatively probe this scenario. Using Eq. 3 of \citet[][or Eq. 8 of \citealt{sun2024recurringtidaldisruptionevents}]{Pfahl2005ApJHillsMechanism} the semi-major axis of the captured star can be associated with the properties of the binary (the binary semi-major axis , $a_b$, and the total mass of the binary, $M=M_1+M_2$). Assuming two solar-like stars and $M_{BH}=10^6 M_\odot$, we find that using $\Delta t$ as the observed period yields $a_b$ values that are between $\sim1.4 R_\star$ to $\sim33.4 R_\star$. We can access if the two stars are far enough to avoid a common envelope by calculating the effective radius of the Roche lobe , $R_{RL}$, using Eq. 2 in \citet{Eggleton1983ApJRocheLobe}. For two cases, namely AT 2022exr and AT 2021uvz (which have a double-peaked light curve morphology), we find that this scenario is not plausible, since $R_{RL}>a_b/R_\star$. To further probe the parameter space of the toy model, we use an $M_{BH}$ range [$10^5,\ 10^8$] with a step of 0.5 dex, as well as unequal mass and radius stellar binaries. For each case we repeat the above analysis and we also consider only the cases where $R_t > R_{Schwarzschild}$. We find that all the multiple flares could be explained by our toy model in the aforementioned parameter space, and hence we cannot exclude the pTDE scenario for any of them.

\begin{table*}
\centering
\caption{\label{tab:repeating}Properties of the repeating flare sample.}
\begin{tabular}{lccccccc}
\toprule
Name & $\Delta t$ (days) & $N_{flares}$ & pTDE  & double TDE & separate events (p) & Other & References \\[3pt] 
\midrule
AT 2024pvu  & $\sim6535$ &    2     & $\checkmark$ &$\checkmark$& $\checkmark$ ($5.7\times10^{-4}$)&  -  &i\\
AT 2022exr  & $\sim110$ ($\sim35$)  &3&$\checkmark$&   ?        & X ($9.6\times10^{-6}$)           &  -  &i\\
AT 2022dbl  & $\sim695$  &    2     & $\checkmark$ &$\checkmark$& X ($6.1\times10^{-5}$)           &  -  &ii, iii\\
AT 2022agi  & $\sim3760$ &    2     & $\checkmark$ &$\checkmark$& $\checkmark$ ($3.3\times10^{-4}$)&  -  &iv\\
AT 2021uvz  & $\sim 55$  &    2     & $\checkmark$ &   ?        & X ($4.8\times10^{-6}$)           &  -  &i\\
AT 2021mhg  & $\sim710$  &    2     & X &   X        & X ($6.2\times10^{-5}$)           & $\alpha$ &v, vi\\
AT 2020vdq  & $\sim995$  &    2     & $\checkmark$ &$\checkmark$& X ($8.7\times10^{-5}$)           &  -  &vii, viii, ix\\
AT 2020acka & $\sim300$  &    2     & $\checkmark$ &$\checkmark$& X ($2.6\times10^{-5}$)           &  $\beta$  &x, xi\\
AT 2019azh  & $\sim5295$ &    2     & $\checkmark$ &$\checkmark$& $\checkmark$ ($4.6\times10^{-4}$)&  -  &xii, xiii\\
AT 2019ehz  & $\sim300$  &    2     & $\checkmark$ &$\checkmark$& X ($2.6\times10^{-5}$)           &  $\beta$  &vii, xi, xiii\\
AT 2018fyk  & $\sim1350$ &    2     & $\checkmark$ &$\checkmark$& X ($1.2\times10^{-4}$)           &  -  &xiv, xv\\ 
\bottomrule
\hspace{0.1cm}
\end{tabular}
\tablefoot{Column 1: name of the TDE. Column 2: time separation between the peaks (the parenthesis for AT 2022exr is the time separation between the second and third flares). Column 3: number of flares. Column 4: partial TDE scenario. Column 5: disruption of a stellar binary scenario. Column 6: disruption of separate stars scenario. Column 7: other potential mechanism ($\alpha$ = TDE followed by SN, $\beta$ = accretion disc fluctuation or a sudden influx of excess orbiting material from the disrupted star). Column 8: references, i) This work, ii) \citet{Lin2024AT2022dblPartialTDE}, iii) \citet{Makrygianni2025arXivAT2022dbl}, iv) \citet{sun2024recurringtidaldisruptionevents}, v) \citet{Munoz2023AT2021mhgASTRONOTE}, vi) \citet{Somalwar2023AT2020vdq_however_AT2021mhg_appendix}, vii) \citet{Yao2023sample}, viii) \citet{Charalampopoulos2023AT2020vdq}, ix) \citet{Somalwar2023AT2020vdq_however_AT2021mhg_appendix}, x) \citet{Guo2025ApJREFEREE}, xi) \citet{Zhong2025ApJREFEREE}, xii) \citet{Hinkle2021AT2019azh}, xiii) \citet{Hammerstein2023ZTF1}, xiv) \citet{Wevers2019AT2018fyk}, xv) \citet{Wevers2023AT2018fyk}.} 
\end{table*}

An alternative explanation of the repeating flares could be the disruption of a stellar binary (double TDE), which has been studied through simulations \citep[e.g.][]{Mandel2015DoubleTDEs, Mainetti2016DoubleTDEs}. The difference with the Hills mechanism is that now both stars are disrupted, either in sequence or with a delay, since after the first TDE, the second star is captured in an elliptical orbit around the SMBH. \citet{Mandel2015DoubleTDEs} use a population model for the binaries approaching the SMBH with small impact parameters and through numerical experiments, which follow Newtonian stellar dynamics of binaries that approach a SMBH. They found that 18$\%$ of their simulations yield the subsequent disruption of both stars. As noted by \citet{Mandel2015DoubleTDEs}, this channel could produce a double-peaked flare similar to the ones observed in AT 2022exr and AT 2021uvz, but they stress that hydrodynamical simulations would be needed to test this hypothesis. \citet{Mainetti2016DoubleTDEs} modelled light curves of double TDEs using smoothed particle hydrodynamics simulations; however, these simulations did not produce double peaked light curves. Nevertheless they did find a knee\footnote{Noticeable change in the slope of the light curve.} in the decay of the simulated light curves in the cases where both stars in an equal-mass binary were partially disrupted or in cases with highly unequal-mass binaries. We observe a knee in the decay of the light curve of AT 2019ehz (see Figs. \ref{fig:subfigAT2019ehz_g} and \ref{fig:subfigAT2019ehz_r}), where one of the stars could have been caught in an orbit to explain the second flare $\sim300$ days after the first one. Furthermore, in about $2.5\%$ of their simulations, \citet{Mandel2015DoubleTDEs} found that one star would get fully disrupted, while the second was captured in a bound orbit with a period of at least 6 months (and a median period of roughly 50 years). This scenario could explain the double peaked light curves of all the repeating flare TDEs, except AT 2022exr, AT 2021mhg and AT 2021uvz.

There are also additional mechanisms that might be responsible for the multiple flares. One example is AT 2021mhg, where the TDE flare was followed by a SN Type Ia \citep{Somalwar2023AT2020vdq_however_AT2021mhg_appendix}. Additionally, AT 2019ehz and AT 2020acka display secondary, dimmer flares, which could result from an accretion disc fluctuation or a sudden influx of excess orbiting material from the disrupted star \citep{Guo2025ApJREFEREE}.

\begin{figure*}[htpb!]
\centering
\begin{subfigure}{0.49\textwidth}
	\centering
	\includegraphics[width=\linewidth]{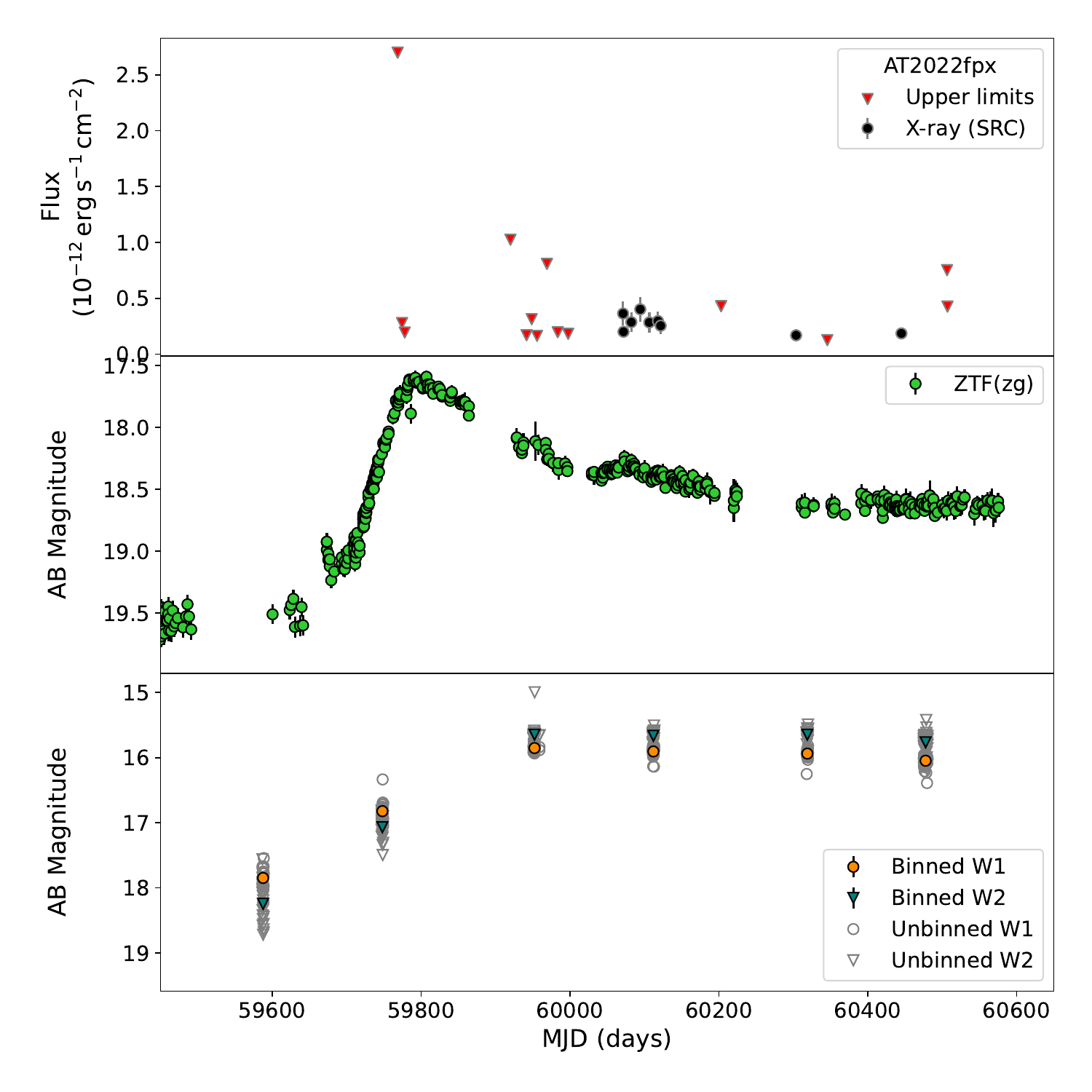}
	\caption{}
	\label{fig:irAT2022fpx}
\end{subfigure}
\hfill
\begin{subfigure}{0.49\textwidth}
	\centering
	\includegraphics[width=\linewidth]{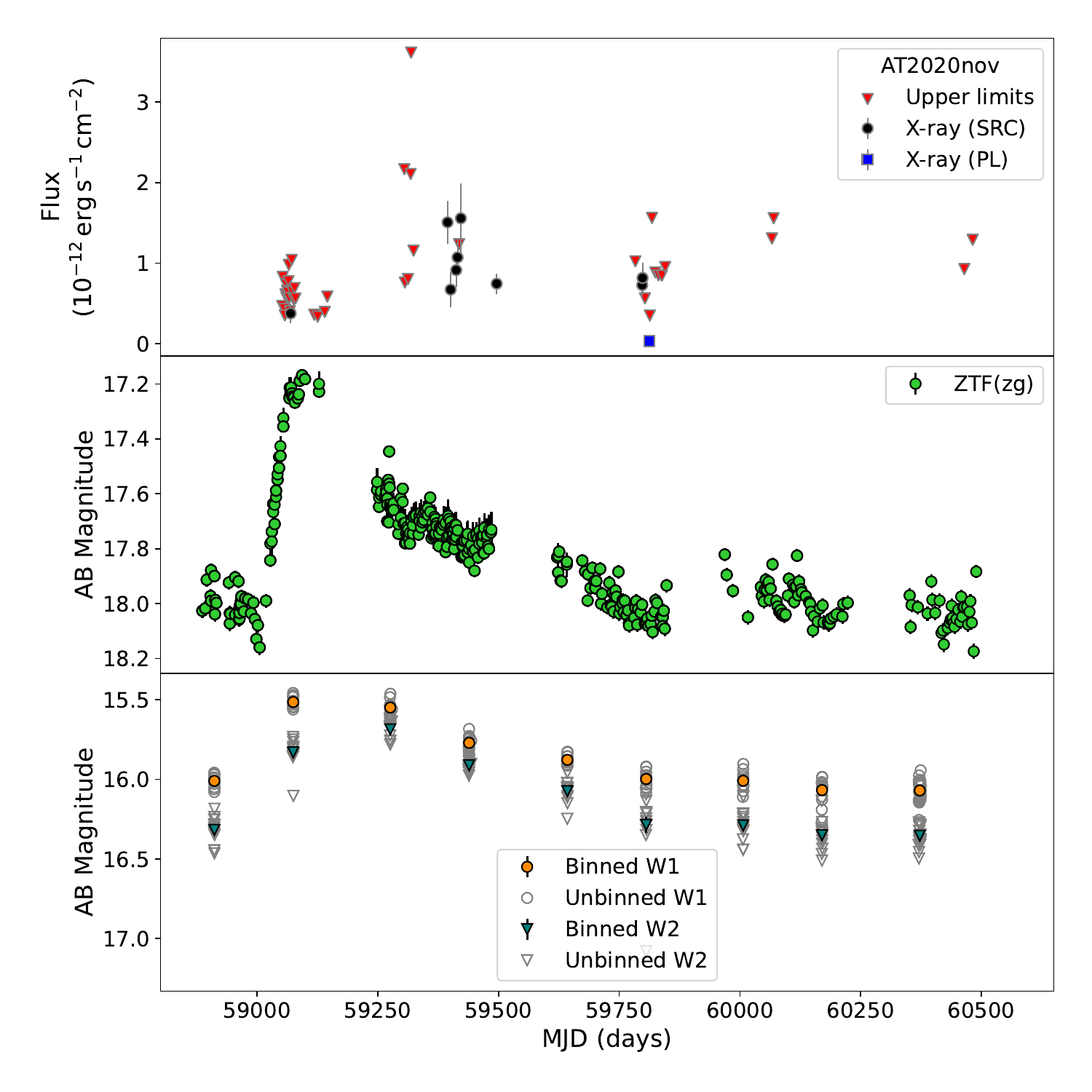}
	\caption{}
	\label{fig:irAT2020nov}
\end{subfigure}
\caption{X-ray, optical, and IR light curves of AT 2022fpx and AT 2020nov in the upper, middle, and bottom panels respectively. For the IR light curves, we show both W1 and W2 binned NEOWISE light curves using orange filled circles and upside-down filled blue triangles respectively, while their un-binned data points are plotted with grey shapes.}
\label{fig:all infrared}
\end{figure*}

\subsection{Multi-wavelength flares}\label{subsect:multiwavelength flares}

A small subset (18) of optical TDEs in our sample display IR flares, namely, AT 2023ugy, AT 2023cvb, AT 2022upj, AT 2022fpx, AT 2022agi, AT 2022dyt, AT 2022aee, AT 2021uqv, AT 2020afhd, AT 2020ksf, AT 2020mot, AT 2020nov, AT 2020pj, AT 2019qiz, AT 2019dsg, AT 2019azh, AT 2017gge, and ASASSN-14li. In Fig. \ref{fig:all infrared} we show two examples of IR light curves (lower panels) from the NEOWISE survey (W1 and W2 filters), in addition to optical (middle panels) and X-ray data (upper panels). Only four TDEs (i.e. AT 2023ugy, AT 2022aee, AT 2020mot, and AT 2020pj) lack an X-ray counterpart (only upper limits are available).

\begin{figure}[htbp!]
\centering
\includegraphics[width=\hsize]{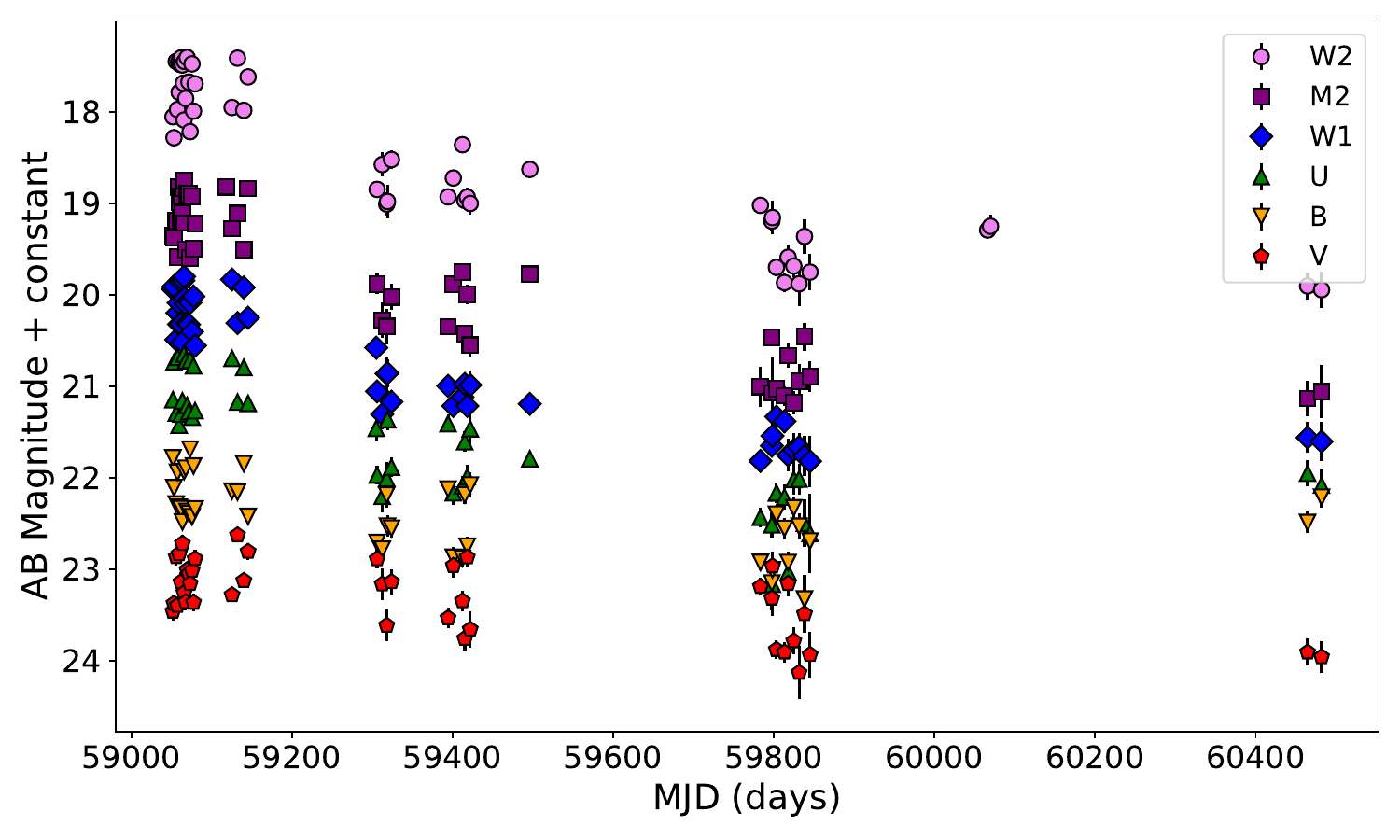}
\caption{UVOT light curves for AT 2020nov.}
\label{fig:uvot example}
\end{figure}

Following the process described in Sect. \ref{sec:uvot}, we retrieved UVOT photometry for 118 TDEs. An example light curve (AT 2020nov) is displayed in Fig. \ref{fig:uvot example}. In total, 111 TDEs have data for all six wavebands, while 95 have at least five observations for each of the three UV bands. These data can be accessed through the GitHub page, under the UVOT folder in the photometry section.

We also investigated TDEs observed in X-rays (see Sect. \ref{subsec:X-ray phot}). In our main sample, 126 sources either have detections or 3$\sigma$ upper limits in the X-rays. Of these, 45 TDEs have at least one detection, while 26 have at least five detections. Example light curves are shown in the upper panels of Fig. \ref{fig:all infrared}. We note that we analyse X-ray observations individually, meaning that stacking observations will lead to more detections. We also consider as detections only the observations with signal-to-noise (S/N) ratio equal to or greater than three. The S/N of the observations is also included in the GitHub files of each source, so the users can set the limit to the threshold they want. All available X-ray, UV and IR data can be found in the accompanying GitHub repository and plotted using the provided app.

\section{Discussion}\label{sect:Discussion}

\subsection{Statistics of the sample}\label{subsect:stat discuss}

In Sect. \ref{subsect:statistics}, we showed that the distributions of the duration, $t_{rise}$ and $t_{decay}$, as well as their ratio, are consistent with a log-normal distribution. These distributions, along with the scatter plot in the bottom-right panel of Fig. \ref{fig:all_distributions}, can be used to generate mock TDE data. We note that this is the first time that the timescales in TDEs have been parametrised with a large sample of more than 100 TDEs. Previously, \citet{vanVelzen2021sample} had a final sample of 39 TDEs, for which they inferred their timescales through fitting, and found that there was no correlation between rise times and decay times. However, \citet{Hammerstein2023ZTF1} and \citet{Yao2023sample}, who studied 30 and 33 TDEs respectively, later reported that there is a positive correlation between rise and decay times, similar to our results (bottom-right panel of Fig. \ref{fig:all_distributions}).

The log-normal behaviour of the redshift distribution could result from observational selection effects, due to the sensitivity of current telescopes and the local universe galaxy density, rather than from some intrinsic property of TDEs. We expect that the distribution of redshifts will change with the addition of data from upcoming surveys such as the Legacy Survey of Space and Time\footnote{\url{https://www.lsst.org/about}} \citep[LSST; ][]{Ivezic2008LSST}. Furthermore, the redshift distribution, along with the rest of the distributions from Sect. \ref{subsect:statistics}, can be used to generate mock TDE populations and light curves.

\subsection{Spectral classes}\label{subsec:spect_class_disc}

As shown in Fig. \ref{fig:pie chart spec}, the majority of TDEs fall into the TDE-H+He category. As predicted by \citet{Syer1999MNRAScitationsREF}, the disruption of main-sequence (MS) stars by SMBHs of mass $\lesssim 10^{8}M_\odot$ would be the easiest for modern-day telescopes to observe. MS stars consist of both hydrogen and helium, hence the observed spectral signatures. Moreover, MS stars are abundant in galactic centres due to their long-lived nature and have been found to dominate TDE rates in past studies \citep[e.g.][]{Kochanek2016MNRAScitationsREF}. The second most common spectral class in our sample is TDE-H, followed by TDE-He and TDE-featureless classes. The statistical study of \citet{Nicholl2022MNRAScitationsREF} uncovered that for the TDE-H sources in their sample, their impact parameter (b; i.e. high b $\rightarrow$ almost full disruption, small b $\rightarrow$ partial disruption) was by far the smallest compared to TDE-H+He and TDE-He, corresponding to partial disruptions. The outer layers of MS stars are mainly hydrogen, so the partial disruption of the outer layers of MS stars could explain the observed TDE-H population. Additionally, hydrogen-rich young MS stars are typically rare, and hence the contribution of their disruption to the population should be small. On the other hand, the TDE-He class is significantly more scarcely observed, with only 16 events (18 if we consider the different epoch classifications from Sect. \ref{subsect:Spectral Type}). This could be explained by the disruption of helium-rich stellar cores, as suggested by \citet{Gezari2012PS1-10jh} for PS1-10jh. These stellar remnants are rare, theorised to be originating from stellar binary systems \citep[e.g.][]{Paczy1967AcAcitationsREF,Iben1985ApJScitationsREF,Gotberg2019A&AcitationsREF}, which could explain their small observed number in TDEs \citep[e.g. ][]{Mockler2024ApJHeliumCores}.

Regarding the TDE-featureless class, we have collected data for 13 such events. \citet{Hammerstein2023ZTF1} found that the 4 TDE-featureless sources in their sample exhibited higher peak bolometric luminosities, hotter temperatures and larger radii than those in the TDE-H and TDE-H+He classes. Moreover, their TDE-featureless sample favoured redder and more massive host galaxies compared to other classes. This trend can be further tested using the TDEs in our catalogue. Additionally, the \citet{Hammerstein2023ZTF1} featureless TDEs were found in relatively higher redshifts than the rest of their sample. Similarly, we find that 7 of our 13 featureless TDEs display redshifts that are 0.25 or above, where the mean is around $0.1$ (median around 0.09). In our sample, $z=0.25$ is at the 91.6th percentile for the optical TDEs (grey filled histogram from Fig. \ref{fig:redshift_distribution}), meaning that 8.4\% of the optical TDEs have $z\ge0.25$. Featureless TDE spectra can arise for several reasons. One possibility is that the line-emitting regions are obscured, depending on our viewing angle. Another explanation is that dense, optically thick outflows reprocess and thermalise high-energy photons, effectively `washing out' sharp emission lines. Additionally, a strong, hot continuum can overpower weaker lines, making them difficult to detect.

To analyse the variability properties of different spectral classes, we applied the same statistical approach used in Sect. \ref{subsect:statistics} to the TDEs for which we could reliably measure flare variability. We then utilised a KS test to assess whether the distributions of different spectral classes are consistent with that of the full sample. We obtained p-values ranging from 0.15 to 0.97 for the rise times, decay times and durations. Although the sample sizes for TDE-H, TDE-He and TDE-featureless classes are small, we find that the timescale distributions of all spectral classes are drawn from the same parent population. \citet{vanVelzen2021sample} had reported that the rise times of their H+He TDEs were longer than the ones for the TDE-H and TDE-He sources. However, this could be due to their smaller sample size along with the small errors from their light curve fitting parameters.

Additionally, we investigated the spectral classes of TDEs in different sub-samples showing either IR emission, repeating flares or X-ray emission. For the IR sample we find that 12 are TDE-H+He, 5 are TDE-H and 1 is TDE-He. For the repeating flare sample we find that 7 are TDE-H+He, 2 are TDE-H and 2 are TDE-featureless. Finally, for the X-ray detection sample we find that 4 are TDE-He, 26 are TDE-H+He, 9 are TDE-H and 3 are TDE-featureless. We estimated the probability that the spectral classes of the different sub-samples are randomly drawn from the full sample using 

\begin{align}
\notag&\mathcal{P}_{IR} = \frac{C^{12}_{74} C^{5}_{23} C^{1}_{13} C^{0}_{12}}{C^{18}_{123}}\approx5.9\times10^{-3}, \\
&\mathcal{P}_{Repeating} = \frac{C^{7}_{79} C^{2}_{23} C^{0}_{16} C^{2}_{13}}{C^{11}_{134}}\approx0.01,\\
&\notag\mathcal{P}_{X\text{-}ray} = \frac{C^{26}_{74} C^{9}_{23} C^{4}_{13} C^{3}_{12}}{C^{42}_{122}}\approx8 \times 10^{-3}
\end{align}

\noindent where $C^k_n$ is the binomial coefficient. Here, the numerator represents the number of ways to compose a sample with the different spectral classes in each sub-sample while the denominator accounts for all possible selections from the full set of 134 TDEs for the repeating flare sample and 123 TDEs for the X-ray and IR samples \footnote{126 TDEs have at least X-ray upper limits; however three of them are jetted TDEs with no optical spectrum available; that is why $k=42$ in the denominator for the X-ray calculation}. The small p-values in all three cases indicate the spectral composition of these sub-samples is unlikely to be random. This could indicate that the spectral characteristics of TDEs are related to their overall multi-wavelength behaviour.

\subsection{Similarities and specific cases in the repeating flare sample}

From our main catalogue, 11 TDEs exhibit more than one flare. While six of these TDEs have been extensively studied in previous works, two of the remaining five sources (AT 2020acka, and AT 2019ehz) have appeared in past sample studies \citep{vanVelzen2021sample, Hammerstein2023ZTF1, Yao2023sample} without their repeating nature being highlighted. Only recently were AT 2020acka's and AT 2019ehz's re-brightening features referenced by \citet{Guo2025ApJREFEREE} and \citet{Zhong2025ApJREFEREE}. Unfortunately, due to a lack of spectroscopic data, we cannot confirm the TDE nature of the flares in the latter half of our repeating flare sample. However, by comparing the rise times, decay times and durations of both flares we find that the timescales for AT 2022dbl, AT 2022agi, AT 2020vdq, AT 2019ehz, and AT 2019azh, are consistent within uncertainty with one another (see Appendix \ref{Appendix:Comparison}). We are unable to compute the rise time of the second flare of AT 2021mhg due to substantial gaps in the respective light curve. Additionally, for AT 2022exr and AT 2021uvz, which are the double peak flares, as well as AT 2020acka, we cannot evaluate the shape of their flares since they are overlapping. From a visual inspection, the rise time of the second flare of AT 2021mhg is clearly different and has been identified as a Type Ia SN. Moreover, the flares in AT 2020acka, possibly originating from accretion disc emission, also displays dissimilar shapes. The similarity of the repeating flares has also been previously noted by \citet{Makrygianni2025arXivAT2022dbl} for AT 2022dbl and \citet{Hinkle2021AT2019azh} for AT 2019azh.

If the additional flares arise from pTDEs, one could expect their shapes to be similar since they would result from the disruption of the same star \citep[simulation papers that have modelled repeating pTDEs have yet to constrain rise times and decay times; e.g. ][]{Liu2025ApJRepeatingREF}. However, the observed diversity in flare shapes suggests that underlying mechanisms may be at play: some flares could indeed be repeating pTDEs, while others may result from distinct events such as double TDEs or unrelated transient phenomena. Moreover, \citet{Liu2025ApJRepeatingREF} found in their simulations that a sun-like star would produce multiple flares of increasing luminosity, consistent with AT 2020vdq, whose second flare is more luminous than the first. However, this picture is not consistent with AT 2022dbl \citep[e.g. ][]{Makrygianni2025arXivAT2022dbl}, where the opposite behaviour is observed.

Furthermore, seven out of the 11 sources in our repeating flare sample exhibit X-ray detections with the exception of AT 2024pvu, AT 2022dbl, AT 2021mhg, and AT 2020acka, for which we obtained only upper limits. We observe that in some TDEs (AT 2022exr, AT 2020vdq, AT 2019ehz, and AT 2018fyk), the X-ray emission follows the optical flare with only a short delay. In contrast, others (AT 2021uvz, AT 2022agi, and AT 2019azh) display significantly longer delays. Notably, for AT 2022exr and AT 2020vdq, X-rays are detected only after the second flare, while for AT 2018fyk they occur nearly simultaneously with the optical outbursts.

\subsection{Possible correlation between X-ray and infrared emission}

As shown in Sect. \ref{subsect:multiwavelength flares}, most TDEs exhibiting IR flares also display X-ray detections. In fact, 14 out of the 18 IR-flaring TDEs have been detected in X-rays. Moreover, in the sample of IR-selected TDEs from the eROSITA sky presented by \citet{Masterson2024IRcandidates}, three of the eight objects (all from their gold sample) were detected in the X-rays at multiple epochs. These findings suggest a potential underlying correlation between IR and X-ray emission.

Notably, both confirmed TDEs associated with astrophysical neutrinos, AT 2019dsg \citep{Stein2021NeutrinoAT2019dsg} and AT 2017gge \citep{Li2024NeutrinoAT2017gge}, are included in our IR flare sample. It has been proposed that neutrino production in these events could be linked to radio emission \citep{Stein2021NeutrinoAT2019dsg}, IR emission \citep{Yuan2024AT2021lwxNeutrino}, or X-ray emission \citep{Li2024NeutrinoAT2017gge}. However, further study is needed to clarify the association between neutrinos and TDEs, as well as the particle acceleration mechanisms that might lead to neutrino production.

We also investigated whether any sources exhibited delayed IR or X-ray emission relative to their optical flares. Our analysis reveals that in two TDEs (AT 2021uqv, AT 2019dsg) only the IR emission is delayed; in another five cases (AT 2022fpx, AT 2020ksf, AT 2020nov, AT 2020afhd, AT 2019azh), only the X-rays are delayed; and in five cases, both IR and X-ray emissions are delayed, with IR flares generally lasting longer, and the X-ray emission being more delayed. In contrast, for AT 2022dyt and ASASSN-14li, the X-ray and IR emission occur almost simultaneously with the optical flare.

Further similarities emerge within our sample. For example, AT 2022fpx, extensively studied by \citet{Koljonen2024AT2022fpx}, displays a light curve similar to that of AT 2020nov, where the X-ray emission coincides with a bump in the optical light curve. Furthermore, all coronal line–emitting TDEs, namely, AT 2022fpx \citep{Koljonen2024AT2022fpx}, AT 2022upj \citep{Newsome2022TNSAT2022upj,Newsome2024AT2022upjCoronal}, AT 2019qiz \citep{Short2023AT2019qizCoronal}, and AT 2017gge \citep{Onori2022AT2017gge}, exhibit both X-ray and IR emission.

The statistical significance of the correlation between IR and X-ray emission in our sample of optical TDEs can be estimated by considering a binomial calculation. The chance that a TDE has X-ray emission can be calculated as the detection fraction of the X-rays in our optical TDE sample (i.e. p=45/126=0.357). We could therefore test the null hypothesis (i.e. that each of the IR-flaring TDEs has an independent chance p=0.357 of X-ray detection) by computing the probability of seeing $k\ge14$ X-ray detections by chance:

\begin{equation}
P=\sum^{18}_{k=14}\binom{18}{k} p^k (1-p)^{18-k} \approx 3.3 \times 10^{-4}
\end{equation}

\noindent Since $P\ll 0.05$, we reject the null hypothesis, meaning that the observed correlation of X-ray and IR emission in our IR-emitting optical TDE sample is likely not by chance. We can also convert the probability to sigma levels for a one-tailed normal distribution, which gives a 3.4$\sigma$ significance. We note that this analysis could be biased towards lower redshift sources, that are brighter and more easily observed by modern day telescopes.

\section{Conclusions}\label{sect:Conclusions}

We have compiled a catalogue of 134 confirmed TDEs up until the end of 2024. We collected multi-wavelength photometry (X-ray, UV, optical, and IR) along with publicly available spectra. The complete dataset is accessible via a dedicated GitHub repository and a Python application. Additionally, we created a list of strong TDE candidates.

We analysed the sources in our main sample by investigating their statistical properties, spectral classes, and the presence of repeating flares, IR emission, and X-ray outbursts. Our key findings are as follows:
\begin{enumerate}
\item By implementing a custom Bayesian Blocks algorithm, we estimated the duration, rise time, decay time, and their ratio for the optical flares of 101 TDEs in our sample. A total of 103 flares were analysed, including all flares from repeating pTDEs (AT 2022dbl and AT 2020vdq). We find that the distributions of the $t_{rise}/t_{decay}$ ratio, durations, rise times, and decay times are well described by a log-normal distribution.
\item The best-fit line for $log_{10}(t_{rise})$ versus $log_{10}(t_{decay})$ is given by $log_{10}(t_{rise}) = \alpha log_{10}(t_{decay})+\beta$, where $\alpha = 0.915 \pm 0.075$ and $\beta = -0.31 \pm 0.17$, with 95\% confidence intervals of 0.768 to 1.062 and -0.65 to -0.04 for $\alpha$ and $\beta$ respectively. This result indicates that the majority of the TDE population is confined within the scatter of this best-fit line. We also find a positive correlation between $log_{10}(t_{rise})$ and $log_{10}(t_{decay})$. These results, in combination with the log-normal distributions of the timescales, can be used to create artificial light curves for future statistical studies.
\item Our spectral analysis revealed that the majority of TDEs belong to the TDE-H+He class, followed by the TDE-H class, with the TDE-He and TDE-featureless occurring at slightly lower frequencies. This distribution is consistent with expectations based on the predominance of main-sequence stars in galactic centres. Although we computed timescale statistics for the different spectral classes, we did not identify any definite trends. 
\item We examined whether specific spectral classes tend to exhibit repeating flares, IR emission, or X-ray outbursts. Our analysis revealed that the spectral class distribution within these TDE sub-samples is unlikely to have arisen by random chance.
\item We identified three new TDEs that show secondary flares in their optical light curves: AT 2024pvu, AT 2022exr, and AT 2021uvz. We further studied recently referenced TDEs displaying re-brightenings: AT 2020acka, and AT 2019ehz. We also compared the shape of the multiple flares for each TDE and found that their shapes are generally similar, excluding AT 2021mhg and AT 2020acka, which exhibit distinct differences.
\item We observed a potential correlation between IR and X-ray emission. Most TDEs (14 out of 18) displaying IR flares also exhibit X-ray detections. We plan to explore this finding further in a future study. Moreover, all reported coronal line emitting TDEs were found to exhibit both X-ray and IR emission.
\end{enumerate}

This comprehensive (mostly optical) TDE catalogue not only provides a robust dataset for statistical studies and machine learning applications but also paves the way for future population studies aimed at understanding the characteristics of still poorly understood TDEs, such as X-ray TDEs \citep{Auchettl2017X-rayCandidateTDEs, Sazonov2021XrayTDE,Khorunzhev2022X-rayTDEs} and IR TDEs \citep{Jiang2021MIRONG,Panagiotou2023IRcandidate,Masterson2024IRcandidates}. As we enter into the `data-rich era' -- especially with forthcoming surveys such as LSST, which are expected to capture tens of TDEs each night -- this work serves as an essential tool for preparing for the challenges and opportunities ahead.

\section*{Data availability}\label{Sect:Data Availability}

Table \ref{tab:honorable} and the catalogue are available in electronic form at the CDS via anonymous ftp to \url{http://cdsarc.u-strasbg.fr/} (130.79.128.5) or via \url{https://cdsarc.cds.unistra.fr/viz-bin/qcat?J/A+A/}. The data and python application (see Sect. \ref{sec:data}), along with the aforementioned table information and overall catalogue, can be accessed through the TDECat GitHub repository\footnote{\url{https://github.com/dlangis/TDECat}}.

\begin{acknowledgements}

The authors thank the anonymous referee for comments and suggestions that helped improve this work. DAL, AP, and IL were funded by the European Union ERC-2022-STG - BOOTES - 101076343. Views and opinions expressed are however those of the authors only and do not necessarily reflect those of the European Union or the European Research Council Executive Agency. Neither the European Union nor the granting authority can be held responsible for them. 

KIIK has received funding from the European Research Council (ERC) under the European Union’s Horizon 2020 research and innovation programme (grant agreement No. 101002352, PI: M. Linares).

NG acknowledge support by a grant from the Simons Foundation (00001470, NG). JD, DM, and ZT took part in this research under the auspices of the Science Internship Program at the University of California Santa Cruz. 

NG is grateful to Caitlyn Nojiri and Hannah Dykaar for their help.

BHTOM has been based on the open-source TOM Toolkit by LCO and has been developed with funding from the European Union’s Horizon 2020 and Horizon Europe research and innovation programmes under grant agreements No 101004719 (OPTICON-RadioNet Pilot, ORP) and No 101131928 (ACME). 

We acknowledge ESA Gaia, DPAC and the Photometric Science Alerts Team (http://gsaweb.ast.cam.ac.uk/alerts).

This work has made use of data from the Asteroid Terrestrial-impact Last Alert System (ATLAS) project. The Asteroid Terrestrial-impact Last Alert System (ATLAS) project is primarily funded to search for near earth asteroids through NASA grants NN12AR55G, 80NSSC18K0284, and 80NSSC18K1575; byproducts of the NEO search include images and catalogues from the survey area. This work was partially funded by Kepler/K2 grant J1944/80NSSC19K0112 and HST GO-15889, and STFC grants ST/T000198/1 and ST/S006109/1. The ATLAS science products have been made possible through the contributions of the University of Hawaii Institute for Astronomy, the Queen’s University Belfast, the Space Telescope Science Institute, the South African Astronomical Observatory, and The Millennium Institute of Astrophysics (MAS), Chile.

The Pan-STARRS1 Surveys (PS1) and the PS1 public science archive have been made possible through contributions by the Institute for Astronomy, the University of Hawaii, the Pan-STARRS Project Office, the Max-Planck Society and its participating institutes, the Max Planck Institute for Astronomy, Heidelberg and the Max Planck Institute for Extraterrestrial Physics, Garching, The Johns Hopkins University, Durham University, the University of Edinburgh, the Queen's University Belfast, the Harvard-Smithsonian Center for Astrophysics, the Las Cumbres Observatory Global Telescope Network Incorporated, the National Central University of Taiwan, the Space Telescope Science Institute, the National Aeronautics and Space Administration under Grant No. NNX08AR22G issued through the Planetary Science Division of the NASA Science Mission Directorate, the National Science Foundation Grant No. AST-1238877, the University of Maryland, Eotvos Lorand University (ELTE), the Los Alamos National Laboratory, and the Gordon and Betty Moore Foundation.

The CSS survey is funded by the National Aeronautics and Space Administration under Grant No. NNG05GF22G issued through the Science Mission Directorate Near-Earth Objects Observations Program.  The CRTS survey is supported by the U.S.~National Science Foundation under grants AST-0909182.

This publication makes use of data products from NEOWISE, which is a project of the Jet Propulsion Laboratory/California Institute of Technology, funded by the National Aeronautics and Space Administration. This publication makes use of data products from the Wide-field Infrared Survey Explorer, which is a joint project of the University of California, Los Angeles, and the Jet Propulsion Laboratory/California Institute of Technology, funded by the National Aeronautics and Space Administration.

This work utilises Matplotlib \citep{hunter2007matplotlib}, NumPy \citep{harris2020array} and SciPy \citep{2020NatMe..17..261V}.

\end{acknowledgements}

\bibliographystyle{aa}
\bibliography{references.bib}

\begin{appendix}
\section{Published TDEs}\label{Appendix:extra_TDEs}

\noindent\textit{AT 2023vto:} This object was initially classified (and still is) as a superluminous super nova Type II (SLSN II) by \citet{Poidevin2023AT2023vto}. This classification was based on faint $H_\beta$ emission that was detected in optical spectra obtained on 2023-11-21 22:30:53 (UTC). However, \citet{Kumar2024AT2023vto} classified it as a TDE, after identifying the broad emission centred at $\lambda 4511\,\text{\r{A}}$ (see their Fig. 4) to be a broad, blueshifted He II $\lambda 4686\,\text{\r{A}}$ emission line (TDE-He; see Sect. \ref{subsect:Spectral Type}).

\noindent\textit{AT 2022cmc:} This is a jetted TDE \citep{Andreoni2022AT2022cmc, Pasham2023AT2022cmc}, similar to Sw J1644 + 57, which unlike the other jetted TDEs, was discovered in optical wavelengths.

\noindent\textit{AT 2022agi:} This is a repeating TDE (see Subsection \ref{subsect:Repeating}), also known as F01004. The first flare was reported in \citet{Tadhunter2017AT2022agi}, while the second was studied in \citet{sun2024recurringtidaldisruptionevents}.

\noindent\textit{AT 2020ksf:} This TDE was first reported in \citet{Gilfanov2020ATelAT2020ksf}, where a soft X-ray transient source was found to be coincident with AT 2020ksf. Furthermore, \citet{Alexander2021AT2020ksf} reported faint radio emission detections coincident with the objects position. 

\noindent\textit{AT 2017gge:} This TDE was first reported in \citet{Wang2022AT2017gge} (ATLAS17jrp) and later appeared in \citet{Onori2022AT2017gge}. Recently, it was reported to be the second TDE associated with a high-energy neutrino in \citet{Li2024NeutrinoAT2017gge}, following AT 2019dsg \citep{Stein2021NeutrinoAT2019dsg}.

\noindent\textit{AT 2017eqx:} This TDE was first reported in \citet{Nicholl2019AT2017eqx}, with broad H {\sc i} and H {\sc ii} emission.

\noindent\textit{LSQ12dyw:} This TDE debuted first in a few circulars \citep{Inserra2012ATel, Smartt2012ATel, Reis2012ATel}, where its nature was discussed. It was later studied and included in the sample of \citet{Charalampopoulos2022LSQ12dyw}.

\noindent\textit{PTF09djl, PTF09ge:} These TDEs were first presented in \citet{Arcavi2014PTFTDEs}, where they were characterised as TDE candidates. Later, they were included in the TDE sample of \citet{Charalampopoulos2022LSQ12dyw}.

\noindent\textit{PS1-10jh:} This TDE first appeared in \citet{Gezari2012PS1-10jh}, where its spectroscopic signature showed broad He {\sc ii} emission lines. It was later included in the \citet{Charalampopoulos2022LSQ12dyw} TDE sample.

\noindent\textit{iPTF15af:} This TDE first appeared in \citet{Blagorodnova2019iPTF15af}, with broad He {\sc ii} emission in its optical spectrum and several other broad features in the UV spectrum. It was later included in the \citet{Charalampopoulos2022LSQ12dyw} TDE sample.

\noindent\textit{iPTF16axa:} This TDE was introduced as a candidate in \citet{Hung2017iPTF16axa}, where it showed broad hydrogen and helium emission lines. It was later included in the \citet{Charalampopoulos2022LSQ12dyw} TDE sample.

\noindent\textit{ASASSN-14li:} This TDE was discovered in December of 2014 \citep{Jose2014ASASSN14li} and has since been studied extensively across the electromagnetic spectrum, in X-rays \citep{Miller2015ASASSN14liXray, Holoien2016ASASSN14liOptXray, Pasham2017ASASSN14liOptXray}, in the optical and near-UV \citep{Cenko2016ASASSN14liUV, Holoien2016ASASSN14liOptXray, Pasham2017ASASSN14liOptXray}, IR \citep{Jiang2016ASASSN14liIR} and radio \citep{Alexander2016ASASSN14liRadio, vanVelzen2016ASASSN14liRadio}. It was later included in the \citet{Charalampopoulos2022LSQ12dyw} TDE sample.

\noindent\textit{ASASSN-14ae:} This TDE was discovered in January of 2014 \citep{Prieto2014ASASSN14aeATel} and was first studied in \citet{Holoien2014ASASSN14ae} as a candidate, where the initial spectrum showed broad hydrogen emission and later evolved into having both broad helium and hydrogen emission. This source was also included in the \cite{Charalampopoulos2022LSQ12dyw} TDE sample.

\noindent\textit{ASASSN-15oi:} This TDE was discovered in August of 2015 \citep{Brimacombe2015ASASSN15oiATel} and was studied in \citet{Holoien2016ASASSN15oi}, where its optical spectrum showed broad helium features. It was later included in the \citet{Charalampopoulos2022LSQ12dyw} TDE sample.

\noindent\textit{Swift J1644+57:} First, this source was thought to be a long lasting $\gamma$-ray outburst, discovered by the {\it Swift} BAT instrument. It was later revealed to be a TDE \citep{Burrows2011SwJ1644, Bloom2011SwJ1644}, since it decayed almost following the TDE characteristic $t^{-5/3}$ power law. Moreover, \citet{Zauderer2011Sw1644Radio} reported radio detections coincident with this source, shortly after its discovery. The properties of this source led to the conclusion that Swift J1644+57 is a highly beamed, non-thermal, relativistic, jetted X-ray TDE.

\noindent\textit{Swift J2058+08:} This source was detected shortly after the discovery of Swift J1644+57, once again by {\it Swift} BAT, sharing many similarities with it. It was first reported in \citet{Krimm2011Sw2058+08ATel} and further studied in \citet{Cenko2012Sw2058+08}, where a multi-wavelength follow-up was initiated. \citet{Cenko2012Sw2058+08} detected a radio counterpart to the flare, suggesting that Swift J2058+08 is the second non-thermal (relativistic) jetted X-ray TDE.

\section{X-ray data reduction}\label{Appendix:X-rays}
\subsection{\textit{Swift}-XRT}\label{sec:swift}

\subsubsection{Data reduction}\label{sec:swift_red}

The XRT photon counting (PC) mode data were downloaded from HEASARC\footnote{\href{https://heasarc.gsfc.nasa.gov/}{https://heasarc.gsfc.nasa.gov/}} data archive, and processed using the \textsc{XRTDAS} software \citep{capalbi2005} developed at the ASI Science Data Center and included in the HEAsoft package (v. 6.33.2) distributed by HEASARC, using a procedure similar to that illustrated in \citet{2013ApJS..209....9P}.

For each observation calibrated and cleaned PC mode event files were produced with the \textsc{xrtpipeline} task (ver. 0.13.7), also producing exposure maps for each observation. In addition to the screening criteria used by the standard pipeline processing, we applied a further filter to screen background spikes that can occur when the angle between the pointing direction of the satellite and the bright Earth limb is low. In order to eliminate this so called bright Earth effect, due to the scattered optical light that usually occurs towards the beginning or the end of each orbit, we used the procedure proposed by \citet{2011AA...528A.122P} and \citet{2013AA...551A.142D}. We monitored the count rate on the CCD border and, through the \textsc{xselect} package, we excluded time intervals when the count rate in this region exceeded \(40 \text{ counts/s}\). In addition we selected only time intervals with CCD temperatures less than \(-50\degree\text{ C}\) (instead of the standard limit of \(-47\degree\text{ C}\)) since contamination by dark current and hot pixels, which increase the low energy background, is strongly temperature dependent \citep{2013AA...551A.142D}.

\subsubsection{Source detection and flux estimates}\label{sec:swift_src}

To detect X-ray sources in the \(0.3-10 \text{ keV}\) XRT images, we made use of the \textsc{ximage} detection algorithm \textsc{detect}, which locates the point sources using a sliding-cell method. The average background intensity is estimated in several small square boxes uniformly located within the image. The position and intensity of each detected source are calculated in a box whose size maximises the S/N.

For each XRT-PC observation, we considered as coordinates of the X-ray counterpart to the TDE source the coordinates of the detected XRT-PC source closest to the TDE source coordinates if this happens to lie closer than \(5\arcsec\)\footnote{This radius represents a conservative value that includes systematic astrometric uncertainties of the various telescopes, as well as positional uncertainties related to the source fluxes. We checked that the majority of the detected X-ray sources fall at separations well below this value, with 70\% of the sources at separations $<$ \(1\arcsec\), 85\% of the sources at separations $<$ \(2\arcsec\), and more than 90\% of the sources at separations $<$ \(3\arcsec\).}. If no XRT-PC source was detected closer than \(5\arcsec\) to the TDE source coordinates, we used the TDE source coordinates themselves.

We then evaluated net \(0.3-10 \text{ keV}\) count rates (or their \(3\) \(\sigma\) upper limits) at X-ray counterpart coordinates with the \textsc{sosta} algorithm that, besides the net count rates and the respective uncertainties, yields the statistical significance of each source. In addition, \textsc{sosta} also estimates the optimal extraction radius \(R_{\text{opt}}\) that maximises the S/N of the source. We note that we used count rates produced by \textsc{sosta} rather than those given by \textsc{detect} because the former are in most cases more accurate, since \textsc{detect} uses a global background for the entire image, whereas \textsc{sosta} uses a local background.

In order to get a first estimate of fluxes for TDE sources X-ray counterparts, we extracted appropriate arf and rmf files at each source location, making use of the \textsc{xrtproducts} task. As extraction regions we used circles centred at the X-ray source coordinates and with a radius equal to \(R_{\text{opt}}\). Assuming black body model with a temperature of \(\sim 5 \times {10}^5\) K\footnote{This value has been evaluated from the error weighted average temperature obtained from the spectral fits, see Appendix \ref{sec:xray-spectra}} and an absorption component fixed to the Galactic value \citep{2005AA...440..775K}, we then converted the net 0.3 - 10 keV count rates evaluated earlier in 0.3 - 10 keV observed intrinsic (i.e. unabsorbed) fluxes.

\subsubsection{Spectral extraction}\label{sec:swift_spectra}

To obtain better estimates on the X-ray source fluxes - as well as possible spectra variability - we extracted XRT-PC source spectra for the selected X-ray counterparts to TDE sources.

In general, source spectra - with the corresponding arf and rmf files - were extracted form events with \textsc{xrtproducts} task, using circular regions centred at the X-ray source coordinates with radii equal \(R_{\text{opt}}\), while background spectra were extracted from annuli centred at the X-ray source coordinates, with inner and outer radii equal to \(2 R_{\text{opt}}\) and \(3 R_{\text{opt}}\), respectively, excluding nearby detected X-ray sources.

When the source count rate is above \(0.5 \text{ counts}/\text{s}^{-1}\), the data are significantly affected by pileup in the inner part of the PSF \citep{2005SPIE.5898..360M}. To remove the pile-up contamination, we extract only events contained in an annular region centred on the X-ray source coordinates \citep[][]{2007AA...462..889P}. While the outer radius of the annulus was set at \(R_{\text{opt}}\), the inner radius was determined by comparing the observed profiles with the updated XRT PSF analytical model\footnote{\href{https://www.swift.ac.uk/analysis/xrt/SWIFT-XRT-CALDB-10_v01.pdf}{https://www.swift.ac.uk/analysis/xrt/SWIFT-XRT-CALDB-10\_v01.pdf}}.

\subsection{\textit{Chandra}-ACIS}\label{sec:chandra}

\subsubsection{Data reduction}\label{sec:chandra_red}

\textit{Chandra}-ACIS data were retrieved from the \textit{Chandra} Data Archive\footnote{\href{http://cda.harvard.edu/chaser}{http://cda.harvard.edu/chaser}}, and were processed and analysed with the \textit{Chandra} Interactive Analysis of Observations \citep[CIAO,][]{2006SPIE.6270E..1VF} data analysis system version 4.16 and Chandra calibration database CALDB version 4.11.1, adopting standard procedures. We run the ACIS level 2 processing with \textsc{chandra\_repro} to apply up-to-date calibrations (CTI correction, ACIS gain, bad pixels), and then excluded time intervals of background flares exceeding \(3\) \(\sigma\) with the \textsc{deflare} task. We produced \(0.3-7 \text{ keV}\) full-band exposure maps, psf maps, and pileup maps with the \textsc{pileup\_map} task.

\subsubsection{Source detection and flux estimates}\label{sec:chandra_src}

We run the \textsc{wavdetect} task to identify point sources in each observation with a \(\sqrt{2}\) sequence of wavelet scales (i.e., 1 1.41 2 2.83 4 5.66 8 11.31 16 pixels) and a false-positive probability threshold of \({10}^{-6}\). 

As we did for \textit{Swift}-XRT (see Sect. \ref{sec:swift_src}), we considered as coordinates of the X-ray counterpart to the TDE source the coordinates of the detected \textit{Chandra}-ACIS source closest to the TED source coordinates if this happens to lie closer than \(5\arcsec\). If no \textit{Chandra}-ACIS source was detected closer than \(5\arcsec\) to the TDE source coordinates, we extracted counts at the TDE source coordinates.

To evaluate the source fluxes we made use of the \textsc{srcflux} task, extracting counts from circular regions centered at the source location, with radii equal to the 99\% EEF PSF radius \(R_{99}\) (as estimated from the PSF maps), while for the background regions we considered annuli with inner and outer radii equal to \(2 R_{99}\) and  \(3 R_{99}\), respectively, excluding nearby detected X-ray sources. Appropriate arf and rmf files were generated for the considered regions, and the net \(0.3-7 \text{ keV}\) count rates were converted to \(0.3-10 \text{ keV}\) observed intrinsic fluxes assuming a model comprising an absorption component fixed to the Galactic value and a black body with a temperature of \(\sim {10}^6 \text{ K}\). In the case of unconstrained count rates, \(3\) \(\sigma\) upper limits were evaluated with the \textsc{aprates\_tool} task, and converted to intrinsic flux upper limits with the \textsc{modelflux} task, adopting the same model described before.

\subsubsection{Spectral extraction}\label{sec:chandra_spectra}

Source spectra and the corresponding arf and rmf files were extracted form event files with \textsc{specextract} task, using circular regions centred at the X-ray source coordinates with radii equal \(R_{99}\), while background spectra were extracted from annuli centred at the X-ray source coordinates, with inner and outer radii equal to \(2 R_{99}\) and \(3 R_{99}\), respectively, excluding nearby detected X-ray sources. Also in this case we excluded inner pixels with pileup larger than \(5\%\) as estimated from the pileup maps.

\subsection{\textit{XMM-Newton}-EPIC}\label{sec:xmm}

\subsubsection{Data reduction}\label{sec:xmm_red}

\textit{XMM-Newton}-EPIC data were retrieved from the \textit{XMM-Newton} Science Archive\footnote{\href{http://nxsa.esac.esa.int/nxsa-web}{http://nxsa.esac.esa.int/nxsa-web}} and reduced with the \textsc{SAS}\footnote{\href{http://www.cosmos.esa.int/web/xmm-newton/sas}{http://www.cosmos.esa.int/web/xmm-newton/sas}} 21.0.0 software.

Following \citet{2005ApJ...629..172N} we filtered EPIC data for hard-band flares by excluding the time intervals where the \(9.5-12 \text{ keV}\) (for MOS1 and MOS2) or \(10-12 \text{ keV}\) (for PN) count rate evaluated on the whole detector FOV was more than 3\(\sigma\) away from its average value. To achieve a tighter filtering of background flares, we iteratively repeated this process two more times, re-evaluating the average hard-band count rate and excluding time intervals away more than \(3\sigma\) from this value. The same procedure was applied to soft \(1-5 \text{ keV}\) band restricting the analysis to an annulus with inner and outer radii of \(12\arcmin\) and \(14\arcmin\) excluding detected X-ray sources in the field, where the detected emission is expected to be dominated by the background.

\subsubsection{Source detection and flux estimates}\label{sec:xmm_src}

When possible, we merged \(0.3-10 \text{ keV}\) data from MOS1, MOS2 and PN detectors using the \textsc{emosaic} task, in order to detect the fainter sources that would not be detected otherwise. Sources were detected following the standard SAS sliding box task \textsc{edetect\_chain} that mainly consist of three steps: 1) source detection with local background, with a minimum detection likelihood of 8; 2) remove sources in step 1 and create a smooth background maps by fitting a 2-D spline to the residual image; 3) source detection with the background map produced in step 2 with a minimum detection likelihood of 10. The task \textsc{emldetect} was then used to determine the parameters for each input source - including the count rate - by means of a maximum likelihood fit to the input images, selecting sources with a minimum detection likelihood of 15 and a flux in the \(0.3-10 \text{ keV}\) band larger than \({10}^{-14}\text{ erg}\text{ cm}^{-2}\text{ s}^{-1}\) (assuming an energy conversion factor of \(1.2\times {10}^{-11}\text{ cts}\text{ cm}^{2}\text{ erg}^{-1}\)). An analytical model of the PSF was evaluated at the source position and normalised to the source brightness. The source extent \(R_{\text{ext}}\) was then evaluated as the radius at which the PSF level equals half of local background.

Again, we considered as coordinates of the X-ray counterpart to the TDE source the coordinates of the detected \textit{XMM-Newton}-EPIC source closest to the TDE source coordinates if this happens to lie closer than \(5\arcsec\). If no \textit{XMM-Newton}-EPIC source was detected closer than \(5\arcsec\) to the TDE source coordinates, we considered as coordinates of the possible X-ray counterpart the TDE source coordinates themselves. In this case, the source \(0.3-10 \text{ keV}\) net count rate (or its \(3\) \(\sigma\) upper limits) and optimal extraction radius (maximising the signal to noise ratio of the source) was provided by \textsc{eregionanalyse}.

Again, to estimate the fluxes of TDE sources X-ray counterparts, we extracted appropriate arf and rmf files for each available detector at each source location, making use of the \textsc{arfgen} and \textsc{rmfgen} tasks. As extraction regions we used circles centred at the X-ray source coordinates and with a radius equal to \(R_{\text{ext}}\). Assuming a black body model with a temperature of \(\sim {10}^6 \text{ K}\) and a absorption component fixed to the Galactic value, we then converted the net \(0.3-10 \text{ keV}\) count rates of each detector \(0.3-10 \text{ keV}\) intrinsic (i.e. unabsorbed) fluxes. The mean source flux was then finally evaluated as the mean of the fluxes of each available detector weighted by its uncertainty, and the uncertainty on this mean flux was evaluated as
\begin{equation}
	\bar{\sigma} = \sqrt{1/\sum_i{\left({1/\sigma_i^2}\right)}},
\end{equation}
where the sum runs on the available detectors \citep{2020AA...641A.136W}.

\subsubsection{Spectral extraction}\label{sec:xmm_spectra}

The source spectra were extracted with the \textsc{evselect} task from circular regions centred at the X-ray source coordinates with radii equal \(R_{\text{ext}}\), while background spectra were extracted from annuli centred at the X-ray source coordinates, with inner and outer radii equal to \(2 R_{\text{ext}}\) and \(3 R_{\text{ext}}\), respectively, excluding nearby detected X-ray sources. The corresponding arf and rmf files were generated with the \textsc{rmfgen} and \textsc{arfgen} tasks to take into account time and position-dependent EPIC responses.
Again, inner regions of high pileup were estimated and excluded using the \textit{epatplot} task through the distortion of pattern distribution, following the procedure explained in the SAS Data Analysis Threads\footnote{\href{https://www.cosmos.esa.int/web/xmm-newton/sas-thread-epatplot}{https://www.cosmos.esa.int/web/xmm-newton/sas-thread-epatplot}}. 

\subsection{X-Ray spectral fitting}\label{sec:xray-spectra}
Spectral fitting was performed with the Sherpa\footnote{\href{http://cxc.harvard.edu/sherpa}{http://cxc.harvard.edu/sherpa}} modelling and fitting application \citep{2001SPIE.4477...76F} in the \(0.3-10\text{ keV}\) energy range for \textit{Swift}-XRT and \textit{XMM-Newton}-EPIC spectra, and in the \(0.3-7\text{ keV}\) energy range for \textit{Chandra}-ACIS spectra, adopting Gehrels weighting \citep{1986ApJ...303..336G}. Source spectra were binned to a minimum of 20 counts/bin to ensure the validity of \(\chi^2\) statistics. In addition, for the EPIC spectra we excluded from the spectral fitting the \(1.45-1.55\text{ keV}\) band due to variable Al K lines, and fitted simultaneously MOS1, MOS2 and PN spectra, if available.

For the spectral fitting we used two different models: (1) a model comprising an absorption component fixed to the Galactic value and a power-law with slope \(a\) and (2) a model comprising an absorption component fixed to the Galactic value and a black body with temperature \(T_{BB}\). In order to select the best fit model we made use of the Akaike information criterion \citep[AIC, see for example][]{2007MNRAS.377L..74L}:
\begin{equation}
	AIC=\chi^2+2k,
\end{equation}
where \(k\) is the number of model parameters, so we selected as the best-fit model the one that provides the lower value of AIC.

\subsection{X-ray catalogues}\label{sec:catalogs}
\subsubsection{XMM catalogue}\label{sec:xmm_cat}

We cross-matched the 4XMM-DR13 catalogue with our catalogue of TDEs using a \(5\arcsec\) search radius. Since the fluxes reported in 4XMM-DR13 are derived from the count rates assuming a model comprising a power-law with slope 1.7 and a Galactic absorption of \(3 \times {10}^{20} \text{ cm}^{-2}\) \citep{2020AA...641A.136W}, for uniformity we converted the \(0.2-12 \text{ keV}\) count rates for each detector to \(0.3-10 \text{ keV}\) observed intrinsic fluxes using PIMMS\footnote{\href{https://heasarc.gsfc.nasa.gov/docs/software/tools/pimms.html}{https://heasarc.gsfc.nasa.gov/docs/software/tools/pimms.html}}, assuming a model comprising an absorption component fixed to the Galactic value and a black body with a temperature of \(\sim {10}^6 \text{ K}\). The fluxes for each detector and their respective uncertainties were there converted in mean flux and uncertainty using the same procedure illustrated in Sect. \ref{sec:xmm_src}.

\subsubsection{EROSITA}\label{sec:erosita}

We cross-matched the eRASS1 catalogue with our catalogue of TDEs using a \(5\arcsec\) search radius
The fluxes reported in eRASS1 are derived from the count rates assuming a model comprising a power-law with slope 2.0 and a Galactic absorption of \(3 \times {10}^{20} \text{ cm}^{-2}\) \citep{2024AA...682A..34M}, and therefore we converted the \(0.2-5 \text{ keV}\) count rates reported in the catalogue to \(0.3-10 \text{ keV}\) observed intrinsic fluxes using PIMMS, again assuming a model comprising an absorption component fixed to the Galactic value and a black body with a temperature of \(\sim {10}^6 \text{ K}\).

For the TDE sources for which we did not find an eRASS1 counterpart, we obtained eRASS1 upper limits\footnote{\href{https://erosita.mpe.mpg.de/dr1/AllSkySurveyData_dr1/UpperLimitServer_dr1/}{https://erosita.mpe.mpg.de/dr1/AllSkySurveyData\_dr1/UpperLimitServer\_dr1/}}. The upper limit eRASS \(0.2-5 \text{ keV}\) fluxes were again converted in \(0.3-10 \text{ keV}\) fluxes using PIMMS\footnote{For the eRASS upper limits in the TDECat GitHub page, we provide the time range covered by the survey, since the exact exposure dates are not provided by the eRASS1 catalogue.}.

\section{Archival repeating flares}\label{Appendix:Repeating}

\subsection{AT 2018fyk}

\begin{figure}[htbp!]
	\centering

	\begin{subfigure}{\linewidth}
		\centering
		\includegraphics[width=\linewidth]{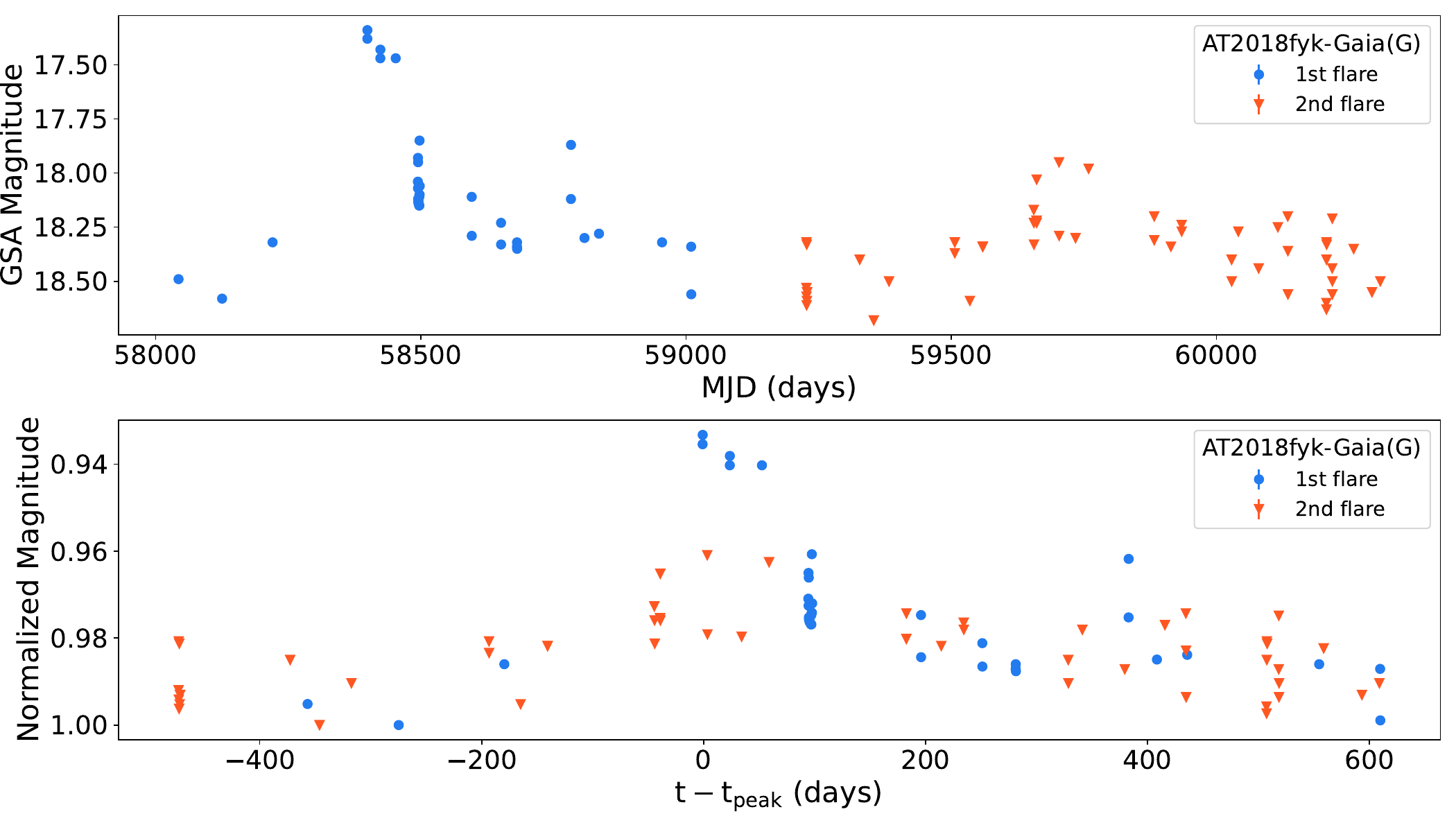}
		\caption{}
		\label{fig:subfigAT2018fyk}
	\end{subfigure}

	\begin{subfigure}{\linewidth}
		\centering
		\includegraphics[width=\linewidth]{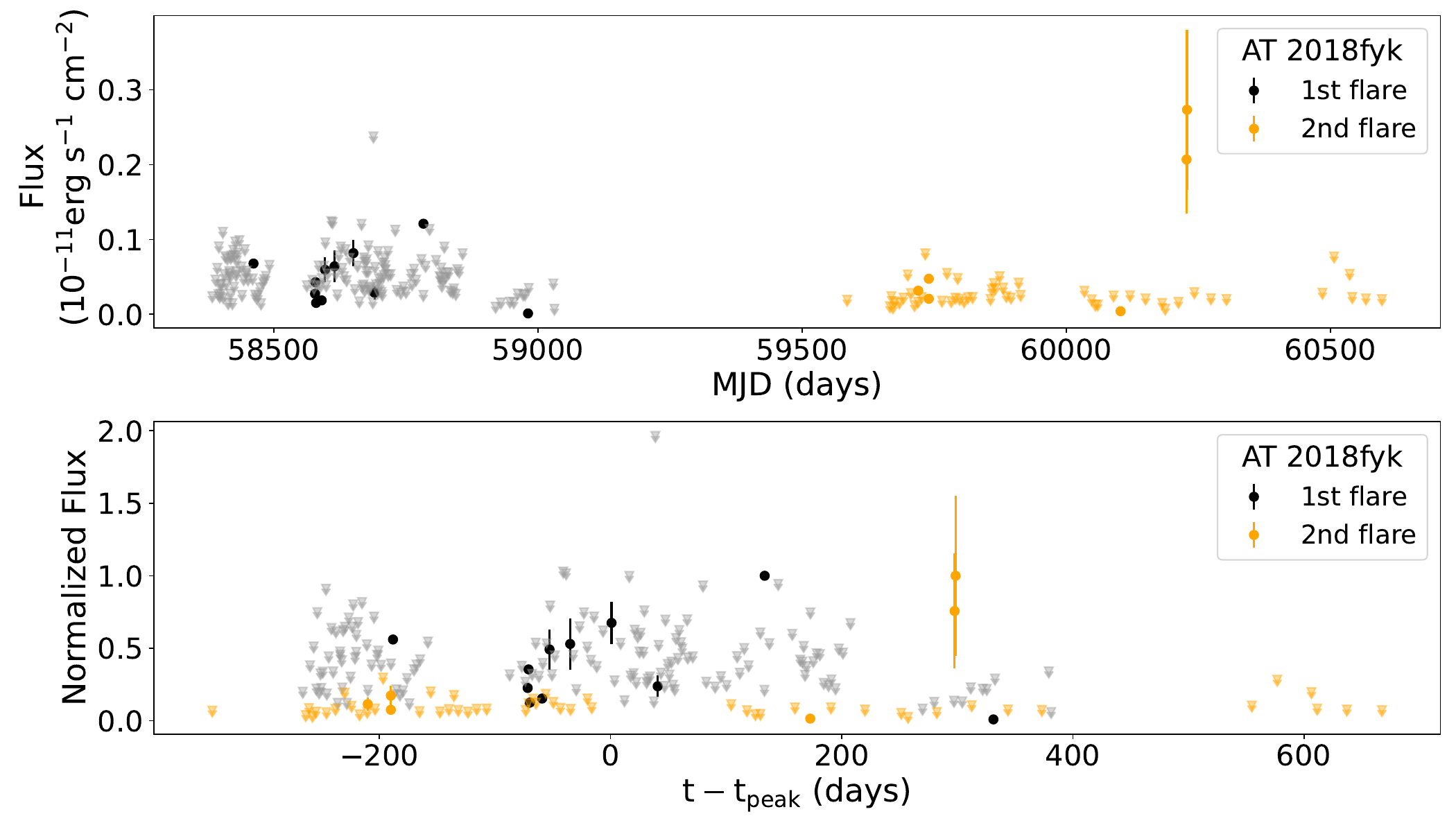}
		\caption{}
		\label{fig:AT2018fyk Xray}
	\end{subfigure}
	
	\caption{Archival optical and X-ray light curves of AT 2018fyk. (a) : Same as for Fig. \ref{fig:alll repeating} but for the Gaia $G$-band light curve of AT 2018fyk. (b) : Same as Fig. \ref{fig:alll repeating} but for the catalogue X-ray data of AT 2018fyk. The black points and orange points show the detections, while the grey and light orange points show the $3\sigma$ upper limits.}
	\label{fig:allAT2018fyk}
\end{figure}

\section{Repeating flare timescale comparison}\label{Appendix:Comparison}

To quantitatively investigate the similarity of the flares in the repeating flare sample, we estimate the rise times, decay times, peak flux and duration (metrics) for the non-confirmed TDE flares as well (i.e. first flare of AT 2022agi, first flare of AT 2019azh, first flare of AT 2024pvu, second flare of AT 2021mhg). We note that we were unable to compute the rise time of AT 2021mhg's second flare due to substantial gaps in the light curves. Additionally, we exclude AT 2018fyk because the second flare is very faint in the optical band. For AT 2022exr and AT 2021uvz, which are the double peak flares, as well as AT 2020acka, we cannot evaluate the shape of their flares since they are overlapping.

\begin{figure}[htbp!]
	\centering
	\includegraphics[width=0.92\hsize]{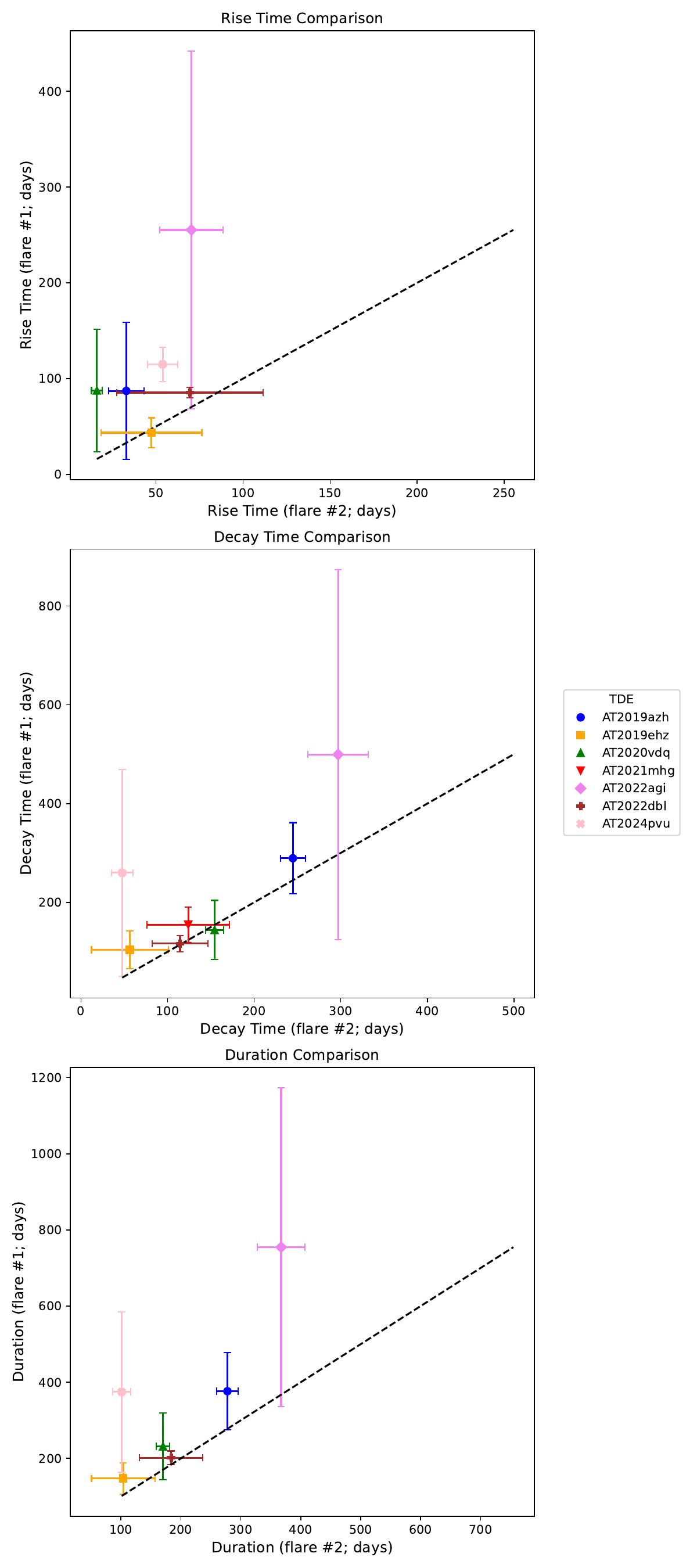}
	\caption{Comparison of the timescales for the first and second flares for seven TDEs in our repeating flare sample.}
	\label{fig:comparison}
\end{figure}

Figure \ref{fig:comparison} shows the metrics of the first flare versus the metrics of the second flare. Almost in all cases (except AT 2024pvu in all timescales), the repeating flares are consistent within uncertainty.

\end{appendix}

\end{document}